\documentclass[fleqn,usenatbib]{mnras}
\usepackage{savesym}
\usepackage{amsmath}
\savesymbol{iint}
\savesymbol{iiint}
\usepackage{txfonts}
\restoresymbol{TXF}{iint}
\restoresymbol{TXF}{iiint}

\usepackage[T1]{fontenc}
\usepackage{ae,aecompl}

\usepackage{graphicx}	% Including figure files
\title[LD of SNe\,Ia ]{Luminosity distributions of Type Ia Supernovae}
\author[C. Ashall]{C.Ashall$^{1}$\thanks{E-mail: c.ashall@2013.ljmu.ac.uk}, P. Mazzali$^{1,2}$, M.Sasdelli$^{1,2}$, S. J. Prentice$^{1}$\\
$^{1}$Astrophysics Research Institute, Liverpool John Moores University, IC2, Liverpool Science Park, 146 Brownlow Hill, \\  Liverpool L3 5RF, UK\\
$^{2}$Max-Planck-Institut f\"ur Astrophysik, Karl-Schwarzschild-Str. 1, D-85748 Garching, Germany}

\begin{document}
\date{Accepted May 2016}

\pagerange{\pageref{firstpage}--\pageref{lastpage}} \pubyear{2017}

\maketitle{}

\label{firstpage}

\begin{abstract}

We have assembled a dataset of 165 low redshift, $z<$0.06, publicly available
type Ia supernovae (SNe\,Ia). We produce maximum light magnitude ($M_{B}$ and 
$M_{V}$) distributions of SNe\,Ia to explore the diversity of parameter space 
that they can fill.  
Before correction for host galaxy extinction we find that the mean $M_{B}$ and
$M_{V}$ of SNe\,Ia are $-18.58\,\pm\,0.07$\,mag and $-18.72\,\pm\,0.05$\,mag
respectively. 
Host galaxy extinction is corrected using a new method based on the SN spectrum. 
After correction, the mean values of $M_{B}$ and 
$M_{V}$ of SNe\,Ia  are $-19.10\,\pm\,0.06$ and $-19.10\,\pm\,0.05$\,mag 
respectively. After correction for host galaxy extinction, `normal' SNe\,Ia
($\Delta m_{15}(B)<1.6$\,mag) fill a larger parameter space in the 
Width-Luminosity Relation (WLR) than previously suggested, and there is 
evidence for luminous SNe\,Ia with large $\Delta m_{15}(B)$. 
We find a bimodal distribution in $\Delta m_{15}(B)$, with a pronounced 
lack of transitional events at $\Delta m_{15}(B)$=1.6\,mag.  
We confirm that faster, low-luminosity SNe tend to come from passive galaxies.
Dividing the sample by host galaxy type, SNe\,Ia from star-forming (S-F) 
galaxies have a mean $M_{B}=-19.20\,\pm\,0.05$\,mag, while SNe\,Ia from passive 
galaxies have a mean $M_{B}=-18.57\,\pm\,0.24$\,mag. 
Even excluding fast declining SNe, `normal' ($M_{B}<-18$\,mag) SNe\,Ia from S-F and passive galaxies are distinct. 
In the $V$-band, there is a difference of 0.4$\,\pm\,$0.13\,mag between the 
median ($M_{V}$) values of the `normal' SN\,Ia population from passive and S-F galaxies. 
This is consistent with ($\sim 15\,\pm\,$10)\% of `normal' SNe\,Ia from S-F 
galaxies coming from an old stellar population. 

\end{abstract}


\section{Introduction}

Type Ia supernovae (SNe\,Ia) are the best standardizable candles in the Universe.
It is currently accepted that SNe\,Ia are produced from a carbon/oxygen White
Dwarf (WD) in a binary system. The three currently favoured progenitor scenarios
are single degenerate (SD), double degenerate (DD) and collisions of C+O WDs in
a triple system. In the SD scenario a C+O WD accretes material from a
non-electron-degenerate companion star \citep{Nomoto97}. In the DD scenario  two
WDs merge after losing orbital angular momentum through the emission of
gravitational waves \citep{WDWD}. Finally, the collision scenario is the head on
collisions of two C+O WDs in a triple system
\citep{Rosswog09,Dong15,KatzDong12,Kushnir13}.  For cosmological uses, Type Ia
supernovae (SNe\,Ia) can be put through intrinsic relations such as the
width-luminosity relation (WLR) and the relation between SN\,Ia colour and
luminosity \citep{Phillips93,mlcs, riess98}. Applying these relations makes
SNe\,Ia standardizable candles, reducing the scatter in their peak luminosity
to $\sim$0.15 mag (e.g. \citet{Jha07,Guy07,conley07}). 

\vspace{5mm}

To first order the luminosity of a SN\,Ia is driven by the amount of radioactive
$^{56}$Ni in the ejecta, and the relation between $^{56}$Ni, luminosity and
opacity determines the LC shape \citep{mazzali07}. However, spectroscopically no
two SNe are exactly identical. They can have different spectral features,
ionizations, velocities and ejecta abundances. There are currently at least 6
subclasses of SNe\,Ia: `normal', 91T-like, 02cx-like, 91bg-like,
`Super-Chandrasekhar' and SNe\,Ia-CSM.  

SN\,91T-like events, named after SN\,1991T, are peculiar and overluminous.
Spectroscopically they are characterised at early times by the weakness or
absence of Ca\,II and Si\,II lines, and show very strong Si\,III lines.
SN\,1991T was $\sim$0.6\,mag more luminous than normal SNe\,Ia
\citep{sasdelli14}. It has been hypothesised that 91T-like events come from
super-Chandrasekhar mass explosion mechanisms \citep{fisher99,Mazzali95}, but
\citet{sasdelli14} finds that a Chandrasekhar mass explosion is more
suited to SN\,1991T. 

SN\,1991bg and SN 91bg-like events have low luminosity, rapidly declining light
curves, low ejecta velocities and a cool spectrum which is dominated by
Intermediate mass elements (IME), as well as particularly strong O\,I and Ti\,II
lines. They are also unusually red and are under-luminous by about $\sim$2\,mag.
The explosion mechanism of a 91bg-like SNe is a matter of debate, with options
including DD explosions of WDs, sub-Chandrasekhar mass explosions triggered by
detonation of the helium layer, or the collision of two WDs
\citep{HilleNiem2000,Hoflich02,Mazzali12,Dong15}.  

SN\,2002cx like events exhibit hot, SN\,1991T-like pre-maximum spectra, a low,
91bg-like luminosity, a LC which is broad for the luminosity but dim for its
shape, and low expansion velocities, roughly half of those of a
normal SNe\,Ia \citep{Li03}.  These events are thought to come from the 
deflagration of a C+O WD, which only experiences partial burning that may 
\citep{Sahu08} or may not \citep{Kromer15} fully disrupt the WD.
 
Super-Chandrasekhar SNe\,Ia are very luminous, which suggests a large
content and therefore a large ejected mass. They are thought to contain $>1.4
M_{\sun}$ of ejecta, and probably come from a DD scenario
\citep{Howell06,Yamanaka09}. However, they can also be explained by an
`interacting scenario', in which a SN\,Ia interacts with a H-/He-poor
circumstellar medium \citep{Hachinger12}. 

Finally, SNe\,Ia-CSM show strong interaction with multiple thin H-rich CSM
shells in the form of H$\alpha\ $ emisson and a black-body-like continuum
\citep{Hamuy03,Deng04,Dilday12,Silverman13}, and are probably SD SN\,Ia.

In modern times there has been a dramatic increase in SN data, which means that
SNe\,Ia can be studied in more detail. One method to do this is by looking at
individual objects, using a time series of spectra. By examining single objects
one can look at the evolution of the abundances in the ejecta and their density
profiles, therefore placing constraints on their explosion properties. This
approach can reveal intrinsic differences between two SNe which can appear to be
photometrically similar, such as SNe\,2011fe and 2014J
\citep{Mazzali13,Ashall14}. 

The increase in available data has also meant that large sample studies can now
be performed,  allowing one to gain more information about the variation in
SN\,Ia properties. One way to separate SNe\,Ia is by host galaxy morphology.
There have been many studies on SNe\,Ia differences depending on host galaxy
type.  It is known that faster declining and intrinsically dimmer SNe\,Ia are
mostly found in passive galaxies \citep{Hamuy95, Hamuy96}.  Peculiar
sub-luminous 91bg-like SNe come from old stellar populations, at least 10 Gyr
old \citep{Howell01}. Host stellar mass was also found to correlate with SN
luminosities; more massive galaxies tend to host SNe\,Ia which have lower
stretch (i.e. a larger rate of decline) than SNe in lower-mass galaxies
\citep{Sullivan10}. The age or metallically of the progenitor may also
contribute to SN\,Ia luminosities, but it is hard to observe these directly.
Studying the early ultraviolet (UV) spectra of SNe\,Ia can let us infer
information about the metallicity of the  progenitor system
\citep{lentz00,sauer,Mazzali13}. It is possible that SNe\,Ia that occur in
galaxies with different star formation, age and dust properties may have  
intrinsically different luminosities \citep{Rigault2013, Childress2013}. It has
also been shown that SNe\,Ia with a higher Si\,II\,6355 line velocity tend to
explode in more massive galaxies \citep{Pan14}. This is not dissimilar from the
result of \citet{Wang13}, who find that SNe\,Ia with high-velocity ejecta are
more concentrated in inner, brighter regions of their host galaxy. 
\citet{Maguire2014} present a comparison of optical spectra with LC width
information from PTF, an untargeted transient survey. They find that on average
SNe\,Ia with a broader LC shape have a larger contribution from the
high-velocity component relative to the photospheric component  in the Ca II NIR
feature.  

A few studies have attempted to build the luminosity functions (LF) of SNe\,Ia.
Such information would be useful because it would make it possible to quantify
the incidence of different subtypes of SNe\,Ia, and thus of their
progenitor/explosion scenarios. \citet{Li2011} present a volume-limited LF from
LOSS, and find evidence for a difference in absolute magnitude between SNe\,Ia
when these are separated into host galaxy bins of E-Sa and Sb-Irr. However, they
do not correct for host galaxy extinction.  \citet{Yasuda10} produce a LF of low
redshift  SN discovered by SDSS-II supernova survey. They claim that the
occurrence of type Ia supernovae does not favour a particular type of galaxy,
but is predominantly dependent on the luminosity of the galaxy. They also claim
that the rate of SNe\,Ia is higher by 31$\pm$35$\%$ in late-type than in
early-type galaxies. \citet{cfa3} present a sample of 185 SNe\,Ia. They find
that 91bg-like SNe\,Ia are distinct from other SNe\,Ia in their colour and light
curve shape-luminosity relation, and state that they should be treated
separately in light curve distance fitter training samples. 

Although the use of SNe\,Ia as cosmological probes is well established, it is
also known that a few events do not follow the normal `rules' of SNe\,Ia. For
example, PTF10ops \citep{PTF10ops} was a sub-luminous SN\,Ia, but its LC had a
normal width. 02cx-like events can also be broad in LC shape and intrinsically
dim.  Traditional LC fitting methods find it difficult to differentiate between
colour and host galaxy extinction, so that, for example, a SN\,Ia which has a
normal LC shape but is intrinsically red maybe mistaken for a SN which is
normal but has more host galaxy extinction. Furthermore, for cosmological
purposes, traditional methods exclude any intrinsically peculiar SNe\,Ia.

The objective of this paper is to include as many peculiar SNe\,Ia as possible,
and hence examine their range and diversity. As there are many subclasses of
SNe\,Ia, we take an approach different from LC template fitting analysis. Our
analysis makes as few assumptions as possible so the intrinsic properties of
these unusual SNe\,Ia can be examined.  This allows for the possibility that SNe
with the same LC shape can have intrinsically different properties. We achieve
this by obtaining the distance to the SNe first, from data that do not use the
SN as distance indicator, rather than obtaining the distance to the SNe from the
LC shape and observed colour information. Although this method is possibly less
accurate, it allows us to explore the full parameter space of SNe\,Ia, and is
better at treating peculiar SNe\,Ia. Furthermore, our method is unique as we
derive host galaxy extinction from the spectra rather than the photometry, which
helps to break the colour-reddening degeneracy \citep{Sasdelli16}, see Section
4. Because we cannot control the observed sample we cannot build a LF, but only
a LD, which is however useful as all SNe used here are nearby, suggesting that
the sample should be reasonably complete.  

In this paper we carry out a large sample analysis, which primarily focuses on
SN\,Ia LC properties, separated by host galaxy type. In Section 2 we discuss the
data and methods used in this paper. Section 3 presents the observed SN\,Ia 
luminosity distributions and discusses the Width Luminosity Relation (WLR). In
Section 4 the data are corrected for host galaxy extinction, and separated by
host galaxy type. In Section 5 we present the full luminosity distribution. In
Section 6 attempts to separate the distributions of SNe\,Ia from young and old
stellar systems. The main discussion is provided in Section 7, and a short
summary is presented in Section 8.

\section{Data \&\ Method}

\begin{figure}
\centering
\includegraphics[scale=0.2]{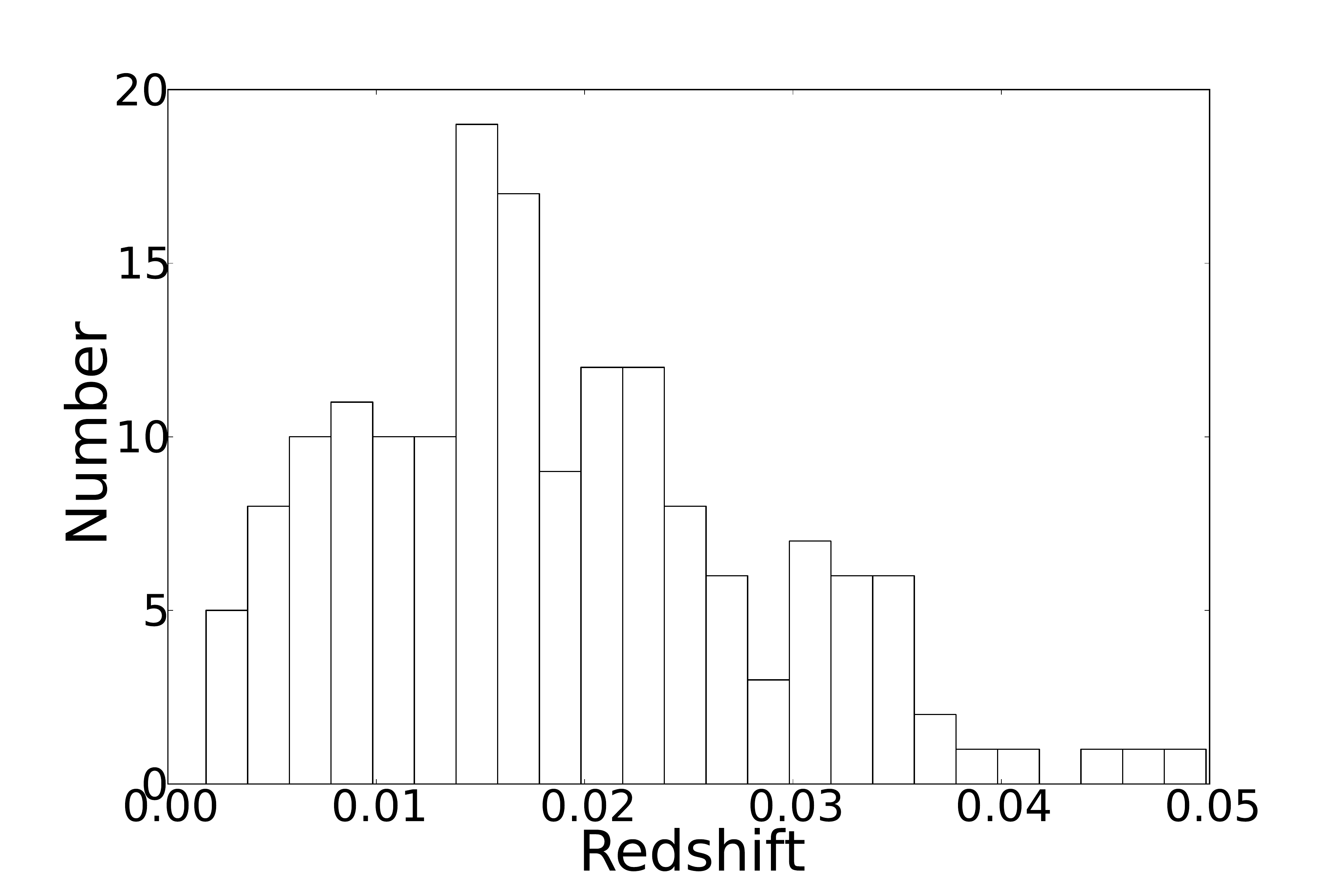}
\caption{Redshift distribution of the SNe\,Ia used in this paper. The bin sizes are z=0.002. }
\label{fig:redsh}
\end{figure}

We have assembled a dataset of 165 low redshift, $z<0.05$, SNe\,Ia with publicly
available data. Their redshift distribution is shown in Figure \ref{fig:redsh}.
The photometric data in this paper come from a variety of public sources,  see
Table 1. The mean redshift of the sample is $z=0.019$. When comparing SN\,Ia LCs
and their derived parameters it is essential that they have sufficient temporal
coverage, as overinterpolating or incorrectly extrapolating the data can cause
incorrect results. All SNe in the data set used in this paper have good temporal
coverage, at least 6 data points, from maximum to +15 days and at least one
pre-maximum data point.  The data used in this paper were published in the
standard Johnsson-Cousin filter system, and no filter conversions were carried
out in this analysis.  We use the $B$ and $V$-band filters for the analysis as
this is where SNe\,Ia peak in flux; these bands are also historically the most
often used, and therefore the best sampled. Most of the optical lines are within
the $B$ and $V$ passbands, therefore these bands are the best to study the
diversity of SNe\,Ia. There is an obvious biases in our sample caused by the
fact that most data were obtained by magnitude-limited surveys. However, we have
used as many SNe\,Ia as possible to avoid small sample statistics. Because the
data come from a range of sources it is not possible to carry out completeness
corrections, therefore in this work we produce luminosity distributions (LD),
which can show the diversity of SNe\,Ia, rather than luminosity functions (LF). 

\begin{table}
 \centering
 \begin{minipage}{65mm}
 \caption{Sources of the data.}
  \begin{tabular}{cc}
  \hline
   Reference &Number of SNe\,Ia \\
   \hline
\citet{Ganeshalingam10}&102\\
\citet{cfa3}&24\\
\citet{cfa4}&18\\
\citet{Reiss99}&10\\
Individual papers\footnote{See appendix A}&11\\
\hline
\end{tabular}
\end{minipage}
\label{table:ref}
\end{table}

In order to avoid introducing any further biases, here we do not assume
that two SNe\,Ia that have the same LC shape necessarily also have the same
intrinsic luminosity and colours \citep[unlike, e.g.,][]{mlcs}. This is done
by first finding an independent distance to each SN and using that distance to
determine the luminosity of the SN. The distances were derived using the local
velocity field model of \citet{2000dist}, which takes into consideration the
influence of the Virgo Cluster, the Great Attractor, the Shapley Supercluster,
and the CMB. To verify the reliability of the distance measurements we checked
against Cepheid distances  for 5 SNe\,Ia which occurred in galaxies with Cepheid
distance measurements. We found that the distances to the SNe\,Ia are consistent
with the Cepheid distances (within 0.06 Mpc). The velocity field model requires
a value of H$_{0}$. We use cosmological parameters which are consistent with
Cepheid measurements, i.e. H$_{0}$=73\,km\,s$^{-1}$\,Mpc$^{-1}$,
$\Omega_{m}=0.27$, $\Omega_{\Lambda}=0.73$. It should be noted that a change in
H$_{0}$ would cause a global shift in values, but it would not directly affect
the results in this paper.

Before different SNe can be compared their photometry must be dereddened and
converted to rest frame. All SNe were corrected for foreground Galactic
extinction using the \citet{newMWred} map and assuming $R_{\rmn{v}}=3.1$.  The
data were converted to the rest frame and K-corrections were applied. We used a
time series of spectra of SN 2011fe \citep{Mazzali13} as a template to calculate
the K-corrections, and carried out the corrections in accordance with
\citet{Oke1968}. Using SN 2011fe as a template is making an assumption about the
SED of the SN, however this affects the fluxes in our final results by less than
5\% in most cases. The K-corrections were applied to each SN at the
corresponding redshift and epoch, using both the $B$ and $V$-bands.

 \begin{table*}
  \centering
 \begin{minipage}{120mm}
 \caption{Statistics of the data, no corrected for host galaxy extinction has been applied.}
 \begin{tabular}{cccccc} 
 \hline 
   & All & active &  passive & E  & S0  \\ 
Amount of SNe &165 &134 &26 &17&9 \\ 
 \hline 
$ \overline{M_{B}}$ &--18.58 $\pm$0.07 &--18.63 $\pm$0.07 &--18.29 $\pm$0.21 &--18.29 $\pm$0.24 &--18.30 $\pm$0.39 \\
$ \sigma{M_{B}}$ &0.82 &0.77 &1.06 &0.99 &1.17 \\
$\overline {\Delta M_{15}B}$ &1.14 $\pm$0.03 &1.11 $\pm$0.03 &1.29 $\pm$0.08 &1.30 $\pm$0.10 &1.28 $\pm$0.12 \\
$\sigma{\Delta M_{15}B}$ &0.32 &0.30 &0.39 &0.40 &0.35 \\
\hline
$ \overline{M_{V}}$ &--18.72 $\pm$0.05 &--18.75 $\pm$0.05 &--18.52 $\pm$0.15 &--18.50 $\pm$0.17 &--18.56 $\pm$0.29 \\
$ \sigma{M_{V}}$ &0.61 &0.58 &0.76 &0.71 &0.86 \\
$ \overline{ \Delta m_{15}V}$ &0.68 $\pm$0.01 &0.67 $\pm$0.01 &0.77 $\pm$0.05 &0.79 $\pm$0.06 &0.74 $\pm$0.07 \\
$ \sigma{ \Delta m_{15}V}$ &0.18 &0.16 &0.25 &0.25 &0.22 \\
 \hline 
$\overline{(B-V)}$\footnote{at $B$ max} &0.11 $\pm$0.02 &0.11 $\pm$0.02 &0.17 $\pm$0.06 &0.16 $\pm$0.07 &0.20 $\pm$0.10 \\ 
$\sigma{(B-V)}^{a}$&0.25 &0.24 &0.29 &0.29 &0.29 \\ 
 \hline 
\end{tabular}
\end{minipage}
\label{table:Absmaghostcorfirst}
\end{table*}

\begin{figure}
\centering
\includegraphics[scale=0.3]{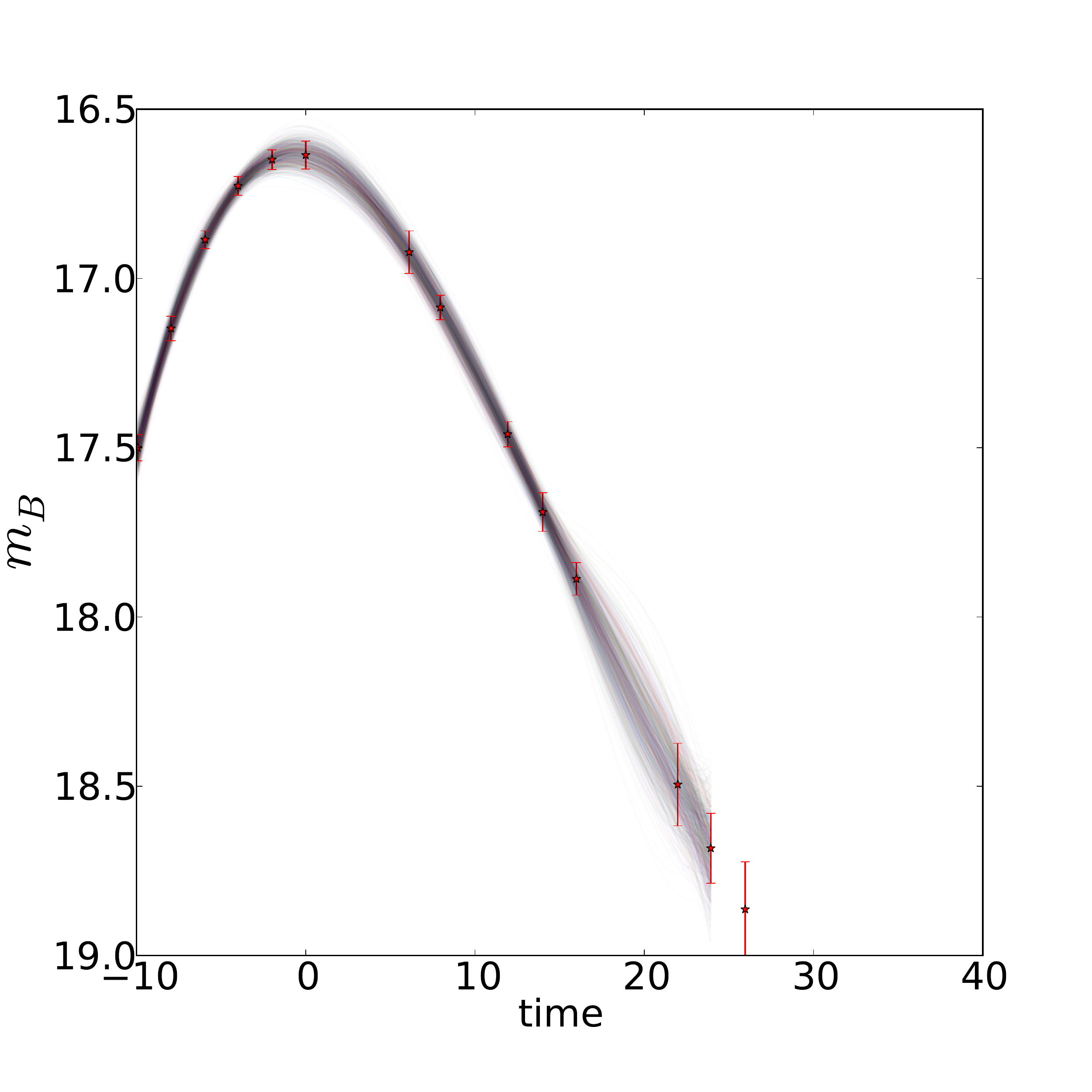}
\caption{An example of the fitting procedure, the $B$-band light curve of SN 2000dn. The red points are the photometry and the shaded area is the 1000 LCs derived from these photometry points, as explained in section 2.1.}
\label{fig:error}
\end{figure}

Finding the host galaxy morphology of each SN in the dataset is integral to our
study. Host galaxy types were obtained from NED\footnote{NASA/IPAC Extragalactic
Database (NED)}. The dataset was separated into two galaxy Hubble type bins,
E-S0 (passive galaxies) and Sa-Irr (star forming (S-F) galaxies). We chose to
separate our sample into these two bins only as the age of the stars is similar
in the galaxies within each bin. Also, using more bins would significantly
decrease the sample in each bin. E+S0 galaxies have an older stellar population
with little or no star formation. Out of our sample of 165 SNe, 134 (82\%) SNe
are from the Hubble Sa-Irr bin, 26 (13\%) are from the E+S0 bin, of which 17
(10\%) are from elliptical galaxies and 9 (5\%) are from S0 galaxies. Finally, 5
SNe (3\%) are from galaxies whose host type could not be determined. Binning the
SNe by by host galaxy type does not decrease the sample as much as
distinguishing by star forming rates or galaxy stellar mass.  

We find that 133 (81\%) of the SNe were classified as spectroscopically
`normal', 14 (9\%) 91T or 99aa-like, 14 (8\%)  at 91bg-like and 4 (2\%) as
02cx-like. Comparing this to the rates of SNe\,Ia from \citet{Li2011} (70\%
normal, 15\% 91bg, 9\% 91T, 5\% 02cx) shows that the public data set has a bias
of too many `normal' SNe\,Ia and fewer dimmer SNe. The lack of 91-bg like events
in our sample could be due to their short rise time, which makes them harder to
detected before maximum light. SNe without maximum light information would be
excluded from the sample in this paper. Additionally, 91bg-like SN are dimmer
events and are therefore affected by Malmquist bias. This does raise the issue
that if one wants a true representation of the intrinsic properties of SNe\,Ia,
a very high cadence, deep and volume-limited survey is required, as well as a
lot of ground-based spectroscopic follow up, however this is not available. It
should be noted that the host galaxy subtraction from \citet{Li2011} excludes
the central $2.4-3.2\,arcsec$ region of the host galaxies, so the true rate may
differ from theirs since they exclude SNe\,Ia from the centre of galaxies.   

SNe\,Ia photometry, to first order, can be analysed using two parameters, the
decline rate or stretch of the LC and the colour correction of the SNe. This is
traditionally done using LC template fitters, and is therefore based on existing
assumptions, data and templates. In reality SNe\,Ia are a far more diverse group
than these LC fitters can assume, as their spectra show. Therefore, we chose to
make no prior assumptions in our LC fitting process, but rather to produce our
own method of fitting the data. In this section we explain how the photometry is
fitted and how we calculate the errors. The LCs are produced by using a smoothed
cubic spline, implementing the {\sc python} module \textsc{univariant spline},
on the photometry which has been K-corrected and dereddened for foreground
Galactic extinction. We use these values as the final values of the apparent
magnitude and light curve shape, as measured by the parameter $\Delta
m_{15}(B)$, the difference in magnitude between maximum and +15\,days
\citep{Phillips93}. This process was carried out for both the $B$ and $V$-bands.
The two spline fits were subtracted to obtain the $B-V$ curve, and from this the
colour at $B$ maximum, $(B-V)_{B\rmn{max}}$, was obtained. The distance modulus
was used to obtain the absolute magnitudes $M_{B}$ and $M_{V}$ of all SNe in the
sample.

To compute the errors on the apparent magnitude and $\Delta m_{15}$ we treat all
of the photometric errors as Gaussians. We randomly vary each photometric point,
in accordance to the weighting of a Gaussian. From this new LCs are produced,
using the method discussed above. This was done 1000 times per SN, and the
standard deviation in the spread of values is taken as the errors on apparent
magnitude and  $\Delta m_{15}(B)$. As an example of this process, Figure
\ref{fig:error} shows the photometry and fitted LCs for SN\,2000dn. The peak
apparent magnitude for SN 2000dn is 16.63$\pm$0.03\,mag and the decline rate
$\Delta m_{15}(B)=1.11\pm 0.07$\,mag. We only fitted the LC up to 30 days past
$B$-band maximum, as we are not analysing late time photometry in this paper.
The plot demonstrates that there is a larger spread in values when the errors on
the photometry are larger. On the other hand, high cadence photometry reduces
the errors. As we obtained SNe from multiple sources we cannot distinguish
between systematic errors which could cause a global shift in the LC, and errors
such as poor host galaxy subtraction, which would affect the shape of the LC.
Therefore the errors of our fits could be overestimated if the errors on the
photometry include systematics.  For more information on the quality of the LC
fit the reader is referred to appendix B.

\section{Luminosity Distribution}
\subsection{$B$ and $V$ luminosity distributions}

\begin{figure*}
\centering
\includegraphics[scale=0.5]{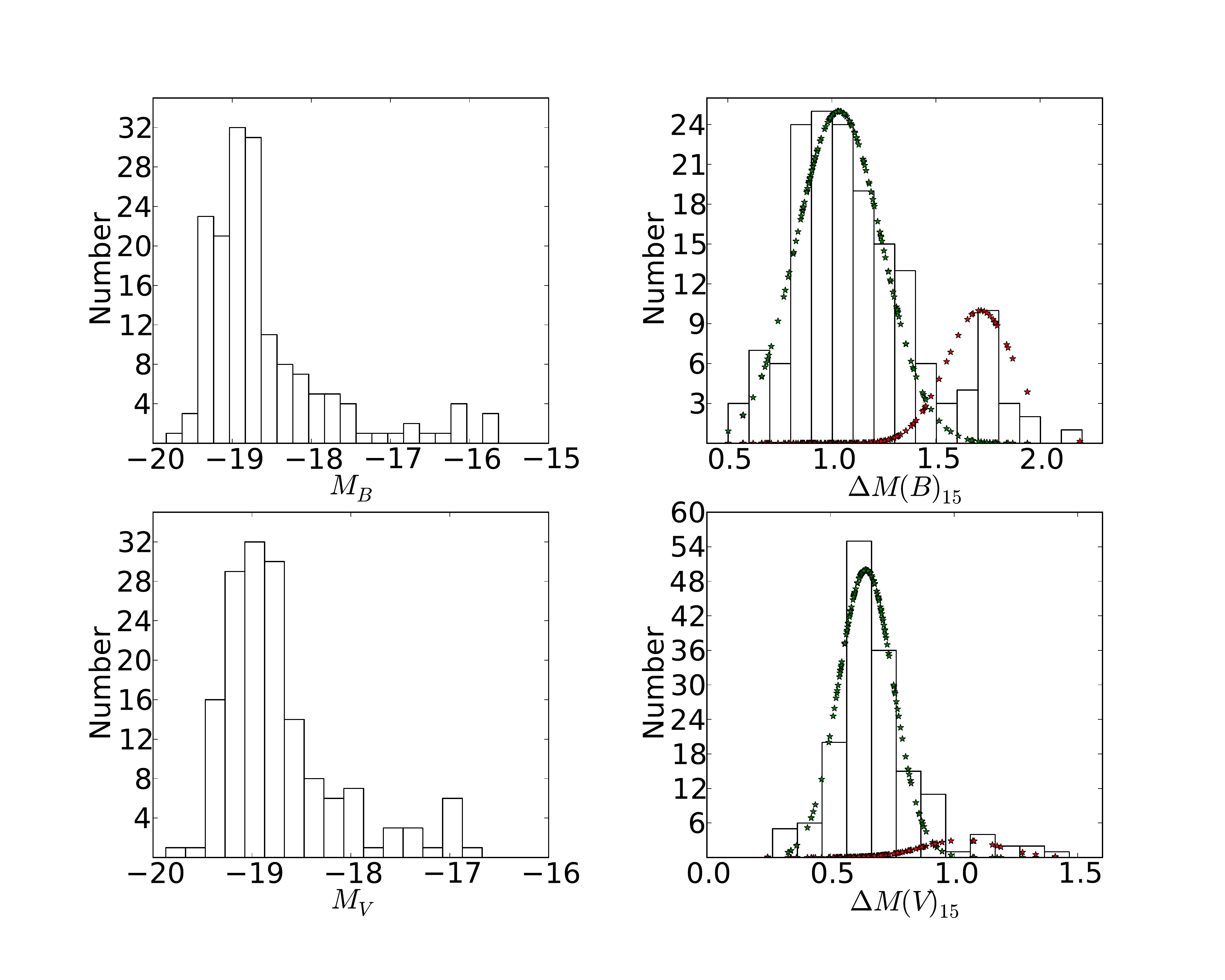}
\caption{The luminosity distributions and $\Delta m_{15}$ distributions of the
SNe sample before correction for reddening. \emph{Top:}  The left plot is a distribution of $B$-band absolute magnitude, and the right plot is the distribution of $\Delta m_{15}(B)$. 
\emph{Bottom:} The left plot is a distribution of $V$-band absolute magnitude, and the right plot is the distribution of $\Delta m_{15}(V)$. In these plots the SNe have not been corrected for host galaxy extinction. The stars in two right hand panels are the Gausians calculated from the GMM tests. }
\label{fig:LDB}
\end{figure*}

The $B$-band absolute magnitude and decline rate distributions of the SNe in our
sample before correction for reddening are shown in the top panels of Figure
\ref{fig:LDB}. The bin sizes are 0.2 for $M_{B}$ and 0.1\,mag for $\Delta
m_{15}(B)$. The mean $B$-band absolute magnitude of the sample is $\bar{M_B}=
-18.58\pm0.07$\,mag, and the sample ranges from $-19.8$ to $-15.6$\,mag. There
is no definitive peak in absolute magnitude in this distribution, which is due
to the missing correction for host galaxy extinction and to the `tail' of
sub-luminous SNe. The distribution in $\Delta m_{15}(B)$ is bimodal, with two
peaks at $0.9-1.0$\,mag and $1.7-1.8$\,mag.
A Gaussian-mixture modelling test \citep[GMM][]{Muratov10} was run to evaluate 
the likelihood that a bimodal distribution is preferred over a unimodal one. It 
is found that for $\Delta m_{15}(B)$ a bimodal distribution is preferred: the 
probability that the $\Delta m_{15}(B)$ distribution is unimodal is less than 
0.1 per cent. The bimodal distributions are assumed to be Gaussians. 
Out of the 165 SNe\,Ia in the sample, the test places $135.3\pm20.8$ SNe (82 
per cent) in a Gaussian with mean decline rate 
$\Delta m_{15}(B)= 1.034 \pm 0.035$\,mag and standard deviation 
0.207$\pm$0.032\,mag, and 29.7$\pm$20.8 SNe (18 per cent) in a Gaussian with 
mean decline rate $1.713\pm0.131$\,mag and standard deviation 
0.164$\pm$0.080\,mag. This is more than the fraction of 91bg-like SNe. 
The bimodal distribution could be due to the presence of sub-Chandrasekhar-mass 
events with large $\Delta m_{15}(B)$ \citep[e.g.\ ][]{Mazzali12,Mazzali11}, or 
to delayed-detonation explosions with a range of transition densities from
deflagration to detonation: \citet{Hoflich02} find that for a range of smoothly 
distributed transition densities there is a lack of SNe\,Ia with $^{56}Ni$ mass 
between 0.15 and 0.25\,$M_{\sun}$. This gap corresponds to the lack of SNe\,Ia at 
$\Delta m_{15}(B)$=1.6\,mag seen in Figure \ref{fig:LDB}.

SNe in the trough between the two peaks are typically classified as
`transitional', with examples being SNe\,2004eo \citep{SN2004eo} and 2012ht
\citep{SN2012ht}. Transitional SNe bridge the gap between `normal' and 91bg-like
SNe\,Ia. The ejected $^{56}$Ni mass (M$_{\rmn{Ni}}$) of transitional SNe\,Ia 
lie near the lower limit of the M$_{\rmn{Ni}}$ distribution of normal SNe\,Ia.
There is still a lack of published literature on transitional SNe\,Ia. It is
these objects which could help define whether 91-bg-like SNe\,Ia arise from a
separate population or there is a smooth distribution of properties SNe\,Ia.
iPTF13ebh, SN\,1986G and SN\,2003hv are all classified as transitional SNe\,Ia. 
None of these events follow the normal paradigm of SNe\,Ia. SN\,2003hv may have 
less mass in the inner layers of the ejecta than Chandrasekhar-mass density
profiles predict: synthetic nebular models were unable to produce the high
Fe\,III/Fe\,II ratio in the optical spectrum along with the low infrared flux 
using a W7 density profile \citep{Leloudas09}, but a lower mass yielded good
results \citep{Mazzali11}. SN\,1986G was the first reported SN of this type. It
was spectroscopically similar to sub-luminous SNe, showing a (weaker) Ti\,II
feature, a large ratio of the Si\,II lines \citep{Nugent05,hanchlinger06} and
narrow unblended lines in the Fe$\sim$4800 \AA\ feature \citep{phillips86G}.
iPTF13ebh showed several strong NIR C\,I lines in the early time spectra, but no
strong C\,I lines in the optical \citep{hsiao13ebh}. 

It should be noted that \citet{Maguire2014} do not find a bimodal distribution
in $\Delta m_{15}$ when they convert their values of stretch to $\Delta m_{15}$.
This may indicate that the double-peak distribution in $\Delta m_{15}$ could be
a selection effect. However, it is unclear how reliable the conversion from
stretch to $\Delta m_{15}(B)$ is. Furthermore, \citet{Maguire2014} use data from
a flux\,limited survey, and therefore the difference between their distribution
and ours could stem from the lack of observed SN 91bg-like events in their
sample. \citet{Burns2014} state that $\Delta m_{15}(B)$ is an unreliable
parameter for fast declining SNe\,Ia with $\Delta m_{15}(B)>1.7$\,mag, because
for these SNe there is a degeneracy between LC shape and $\Delta m_{15}(B)$.
Here, however, $\Delta m_{15}(B)$ is used only to discriminate between normal
and sub-luminous SNe\,Ia, which is sufficient for our analysis. 

The lower part of Figure \ref{fig:LDB} presents the $V$-band absolute magnitude
and $\Delta m_{15}(V)$ distributions. Bin sizes are again 0.2\,mag and 0.1\,mag
respectively. The mean $M_{V}$ value is $\bar{M_V} -18.72\pm$0.05\,mag, while 
$\bar{\Delta m_{15}(V)} = 0.68\pm0.01$\,mag. We find no statistical evidence
that the distribution in $\Delta m_{15}(V)$ is bimodal. Although not
statistically significant, a bimodal $\Delta m_{15}(V)$ distribution has been
plotted in Figure \ref{fig:LDB} (bottom left panel). The two Gaussians have mean
$\Delta m_{15}(V)$ of $0.642\pm0.015$ and $0.999\pm0.201$\,mag and standard
deviations of 0.106$\pm$0.031 and 0.166$\pm$0.067\,mag, respectively. 

The standard deviation in absolute magnitude, see Figure \ref{fig:LDB}, is less in the $V$-band (0.61\,mag),
than in the $B$-band (0.82\,mag). This
is partly caused by the fact that the $B$-band is more affected by extinction.
It could also be caused by some SNe\,Ia having broad, a High Velocity Ca\,II
feature at $\sim$3800\,\AA\, or different amount of line blanketing in the UV,
although the latter will not be the largest contributor to the difference.

\begin{figure}
\centering
\includegraphics[scale=0.2]{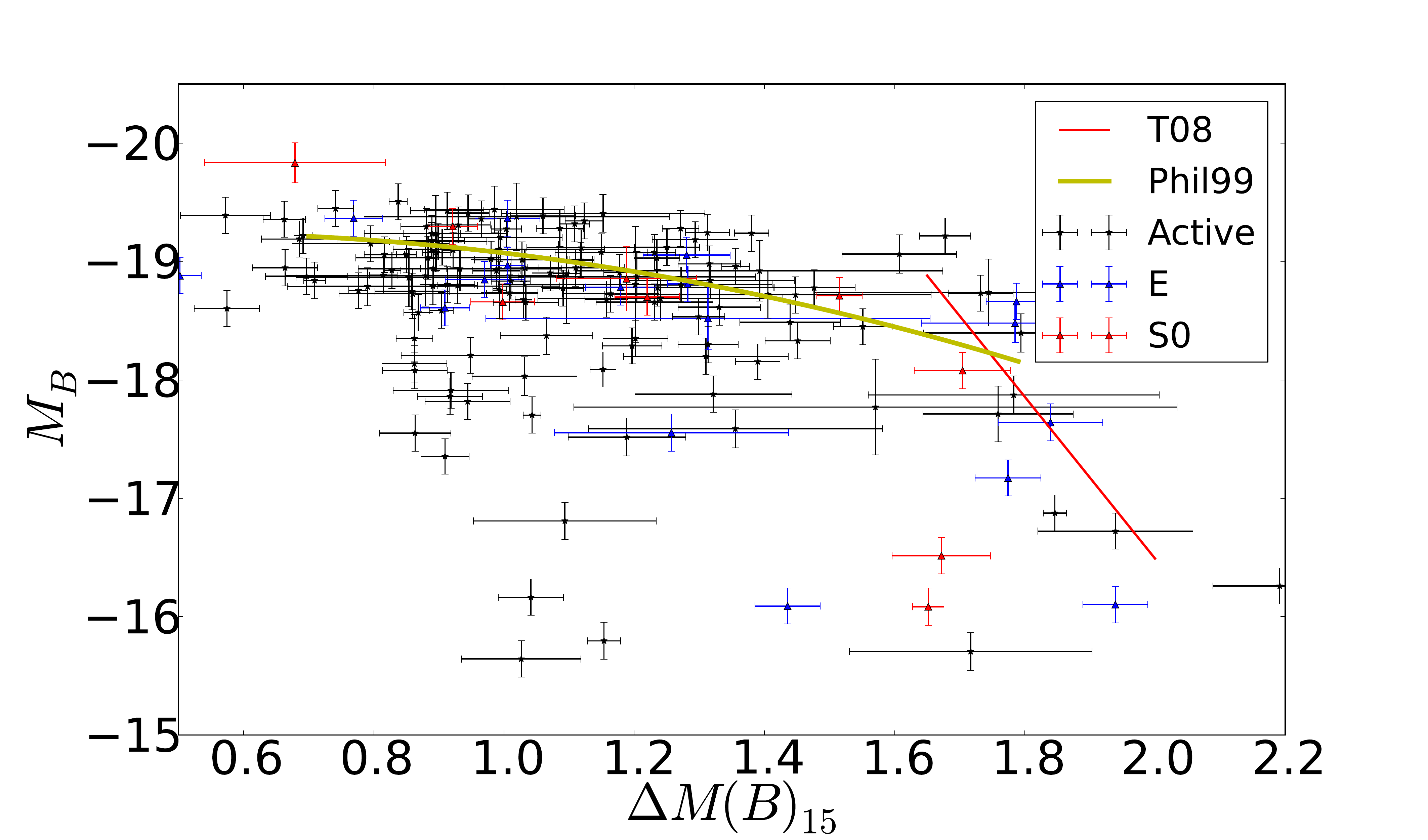}
\caption{The $B$-band WLR before reddening correction. The black points are SNe 
from S-F galaxies, the blue points are SNe from E galaxies, and the red points 
are SNe\,Ia from S0 galaxies.  The yellow line is the WLR given in 
\citet{phillips99}, and the red line is the relation given for the subluminous 
tail from \citet{Taubenberger08}.}
\label{fig:WLRB}
\end{figure}

\subsection{Width Luminosity Relation (WLR)}

Type Ia SNe are known to show an inverse correlation between their absolute
magnitude and the rate at which they decline, $\Delta m_{15}$. This is as the
widths of the LCs are a function of ejecta mass, $^{56}$Ni mass, kinetic energy
and opacity \citep{mazzali07}. The WLR relationship was first published by
\citet{Phillips93}, and extended by \citet{phillips99} and then
\citet{Taubenberger08}. Figure \ref{fig:WLRB} shows this relation within the
context of our work. It is clear that there is an underlying correlation (`the
Phillips Relation') with some scatter. The relation from \citet{phillips99} is
plotted in Figure  \ref{fig:WLRB} to show this. SNe\,Ia with low luminosity but
$\Delta m_{15}(B)<1.6$ are likely to be affected by host galaxy extinction, but
we cannot determine this using only this plot. Faster-declining SNe tend to come
from passive galaxies and are less affected by extinction. Therefore, the `tail'
in the WLR, which was first pointed out by \citet{Taubenberger08}, is likely to
be intrinsic, and not the effect of extinction.  Their relation for the
subluminous tail is also plotted in Figure \ref{fig:WLRB}. Furthermore, there is
a dearth of SNe at $\Delta m_{15}(B) \sim 1.6$\,mag, where the SNe transition
from `normal' to sub-luminous. Interestingly this gap is where the two relations
from \citet{phillips99} and \citet{Taubenberger08} meet. 

\begin{figure}
\centering
\includegraphics[scale=0.2]{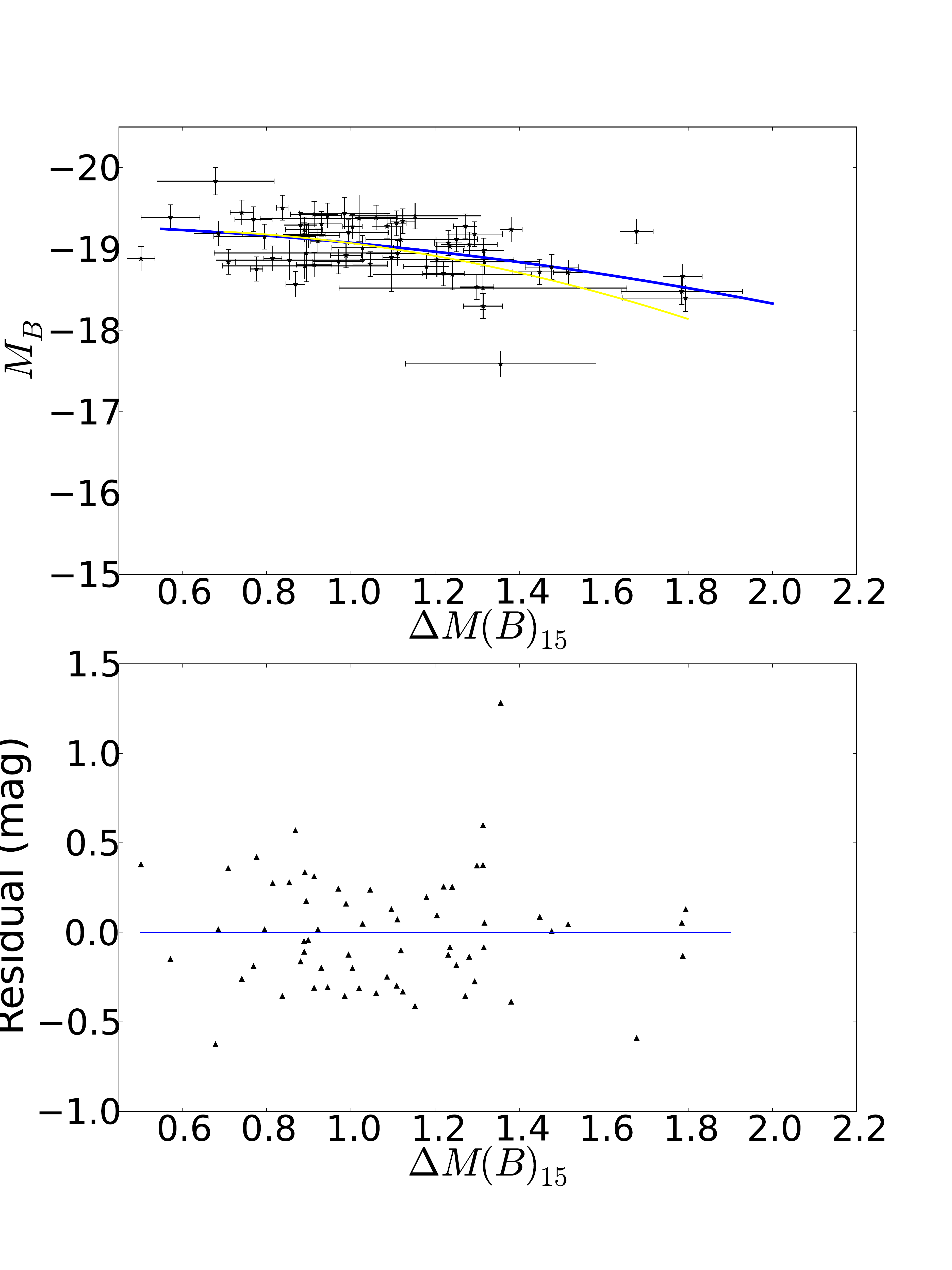}
\caption{\emph{Top:} The WLR before reddening correction including only SNe 
with $(B-V)_{Bmax}<0.01$. The blue line of best fit is a second order 
polynomial. The yellow model is the line of best fit from \citet{phillips99}.
\emph{Bottom:} The residuals of the WLR above as a function of 
$\Delta m_{15}(B)$.  }
\label{fig:WLRLowCB}
\end{figure}

To overcome the effect of extinction, one could select only SNe with a very 
small colour term, $(B-V)_{\rmn{Bmax}}<0.01$, as they are thought to be less
affected by host galaxy extinction. The top panel in Figure \ref{fig:WLRLowCB}
shows that the scatter in the WLR plot is indeed reduced when only these SNe are
selected.  We obtain a second-order polynomial line of best fit, shown as the
blue line in Figure \ref{fig:WLRLowCB}, given by,

\begin{equation} M_{B}=0.252 \Delta m_{15}(B)^2-0.015\Delta m_{15}(B)-19.31 \end{equation}

However, this approach eliminates any intrinsically red or unusual SNe\,Ia. The
bottom plot in Figure \ref{fig:WLRLowCB} shows the residuals from this fit; the
mean residual is 0.22\,mag. The reduced $\chi^{2}$ of the fit is 3.3. When
compared to the fit from \citet{phillips99} (red model in the top panel of
Figure \ref{fig:WLRLowCB}), we find a similar result, although our fit is not as
tight. The SN which sits clearly off the Phillips relation, with a value of
$M_{B}>-18$\,mag is SN 2008cm. SN 2008cm has been rejected in cosmological
studies for being an outlier \citep{2014Rest}. This demonstrates that LCs alone do
not contain enough information to differentiate between types of SNe Ia.

\section{Correction for host galaxy extinction}

The intrinsic colours of SNe\,Ia change over time because of the changing
properties of the ejecta. For less luminous SNe this change is over a shorter
time scale than for more slowly declining SNe. It is observationally difficult
to distinguish between intrinsic colour changes and reddening due to dust
between the SNe and the observer. Determining this is central to the full
understanding of the nature of SNe\,Ia, but it is not trivial. Several  studies
tried to estimate SN\,Ia colour. The Lira law \citep{phillips99} found that the
$B-V$ colours of SNe\,Ia, at all decline rates, evolve in a nearly identical
way, from +30 to +90 days past $V$ maximum. It has also been shown that high-
and normal-velocity (NV) SNe exhibit significant discrepancies in $B-V$ and
$B-R$, but not in other colours \citep{Mandel2014}.

\begin{figure*}
\centering
\includegraphics[scale=0.35]{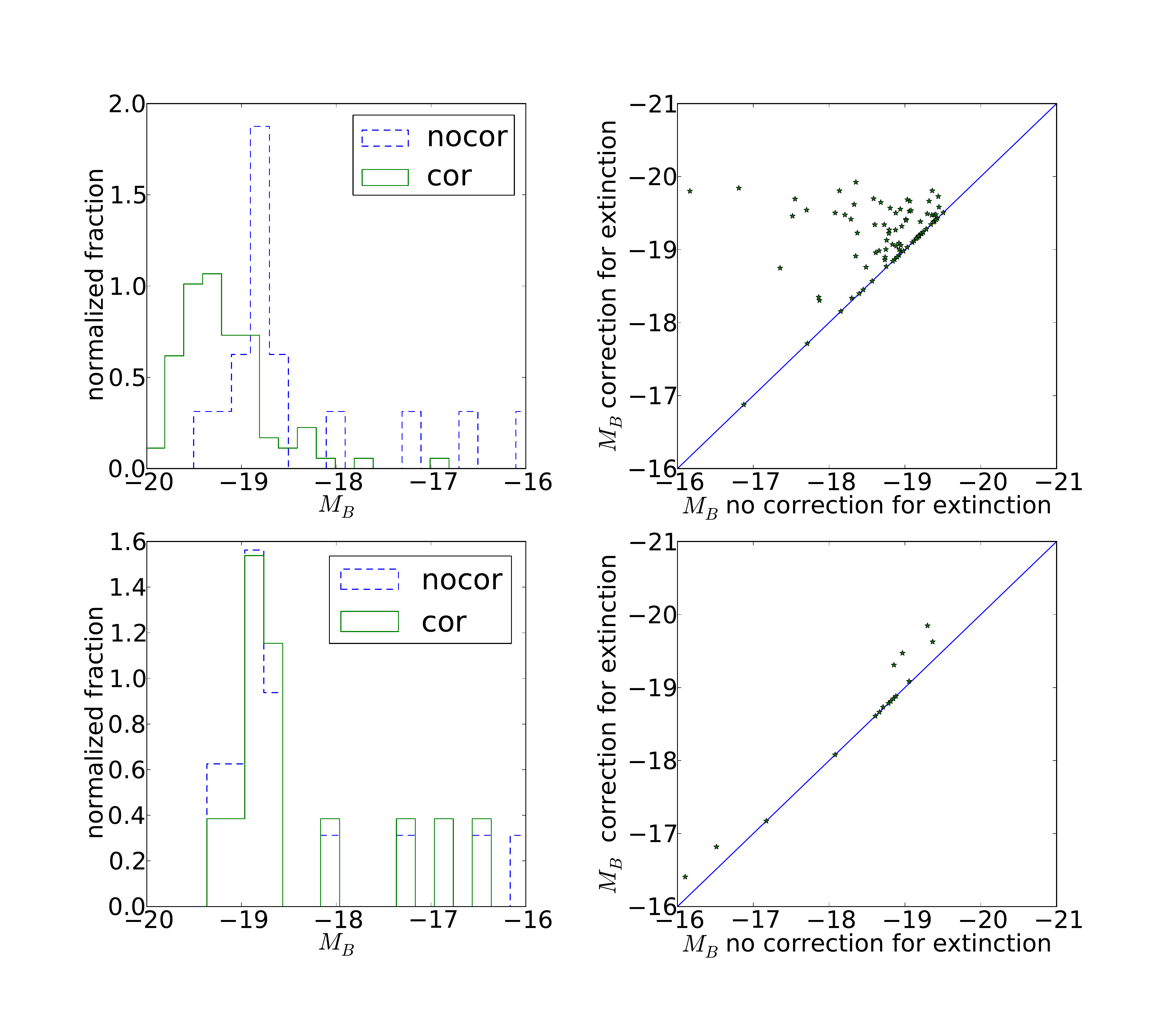}
\caption{\emph{Top:} left: the $M_{B}$ LD of the SN\,Ia from S-F galaxies, 
before and after correction for extinction. 
Right: compares the values of $M_{B}$, again before and after correction for 
extinction, for SNe from S-F galaxies. 
\emph{Bottom:} Left: the $M_{B}$ LD of the SNe\,Ia from passive galaxies, 
before and after correction for extinction. Right: $M_{B}$, before and after 
correction for extinction, for SNe from passive galaxies.  }
\label{fig:hostextinction}
\end{figure*}

Spectral properties such as equivalent widths (EW) can be used as indicators to
attempt to distinguish between colour and host galaxy extinction.  EWs are not affected by
extinction, and there are intrinsic relationships between the EW of certain
lines and the SN LC parameters, such as between EW(Si\,II 5972) and $\Delta
m_{15}(B)$ \citep{Nugent05,hanchlinger06}. When trying to obtain extinction from
estimates of reddening, it is important that the correct value of $R_{\rmn{v}}$
(the total to selective extinction) is used. In the Milky Way the average value
of $R_{\rmn{v}}$ is 3.1, but it varies depending on the region observed. For SNe\,Ia very
low values of $R_{\rmn{v}}$ have sometimes been reported, ranging
from 1.1 to 2.2 \citep{Tripp1998,Kessler2009,guy10,Folatelli10}.
\citet{Chotard2011} derived a larger value of $R_{\rmn{v}}=3.1$, which is
consistent with the MW value \citep{CCMred}. \citet{Sasdelli16} used time-series
of SNe\,Ia spectra in derivative space, which is not affected by poor calibration, to
explore host galaxy extinction, and found that $R_{\rmn{v}}=2.9\pm0.3$ fits the data the
best, and is also consistent with the MW value.

From the WLR in Figure \ref{fig:WLRB}, it is not possible to tell how much of
the scatter is intrinsic and how much is due to extinction, since SNe\,Ia are
not perfectly homogeneous. For almost all astronomical data, correction for both
Galactic and host extinction is important, especially when comparing objects.
\citet{sasdelli14} attempted to break the colour-reddening degeneracy using
spectroscopic time series as predictor variables of the intrinsic colour. They
built a metric space for SNe\,Ia independent of extinction using Principal
Component Analysis (PCA). The intrinsic spectral evolution of the SNe\,Ia is
represented by a 5-dimensional feature space. This space does not include dust
extinction. Two intrinsically similar SNe with different extinction have similar
projections in this feature space. \citet{sasdelli14} use this feature space to
predict the $B-V$ colour of SNe using a Partial Least Square regression (PLS).
Only the intrinsic part of the $B-V$ colour can be predicted by PLS. The
difference between the predicted intrinsic colour and the observed $B-V$ colour
can then be interpreted as due to dust extinction. This yields estimates of 
$E(B-V)$ for the individual SNe. With this host galaxy extinction information we
can examine the intrinsic properties of SNe\,Ia.

In this section we apply a correction for host galaxy extinction based on the
method of \citep{Sasdelli16} and use the values of $E(B-V)$ thus derived.  This
method removes spectroscopically peculiar SNe\,Ia that are underrepresented in
the data, and therefore can bias our sample towards normal and luminous 
SNe\,Ia. We correct the for extinction using the CCM law
\citep{CCMred} with $R_{\rmn{V}}=2.9$, and use the approximation

\begin{equation} \Delta m_{15}(B)_{\rmn{true}}=\Delta m_{15}(B)_{\rmn{obs}}+0.1\times E(B-V),  \end{equation}

to correct $\Delta m_{15}(B)$ for extinction. $\Delta m_{15}(V)$ has not been
corrected for extinction as the effect is negligible in the $V$ band
\citep{phillips99}. 

As the sample size is reduced in this section of the analysis, we first verify
the main properties of the SN sample before and after host galaxy extinction
correction is applied, in order to determine whether the reduced sample is
selected randomly from the larger one or it is biased towards normal SNe\,Ia by
the selection process.  For the sample before correction, 
$\bar{M_B}=-18.66\pm0.07$\,mag, which is comparable to $-18.58\pm\,0.07$\,mag
from the larger uncorrected sample. The mean $\Delta m_{15}(B)$ values from the
two samples are also similar, $1.14\pm0.03$\,mag and $1.11\pm0.03$\,mag for the
larger and smaller samples respectively. The mean $M_{V}$ values were also found
to be similar, $-18.72\pm0.05$\,mag and $-18.77\pm0.05$\,mag for the larger and
smaller samples respectively.
Out of the 56 SNe Ia which were removed from the sample 2 were 91-bg like, 4 were- 91T-like, 
47 were normal SNe Ia. These SNe were rejected as they had insufficient spectral coverage, 
so it is not possible to tell if they have a peculiar spectral evolution. The other 3 of the 56 had 
be to removed from the sample for being spectroscopically peculiar. These were 2005hk 
(a 2002cx like SN), 2004dt (a highly polarised SN, \citep{Altavilla07}) and 1999cl (a highly extinguished SN). 

\begin{figure*}
\centering
\includegraphics[scale=0.5]{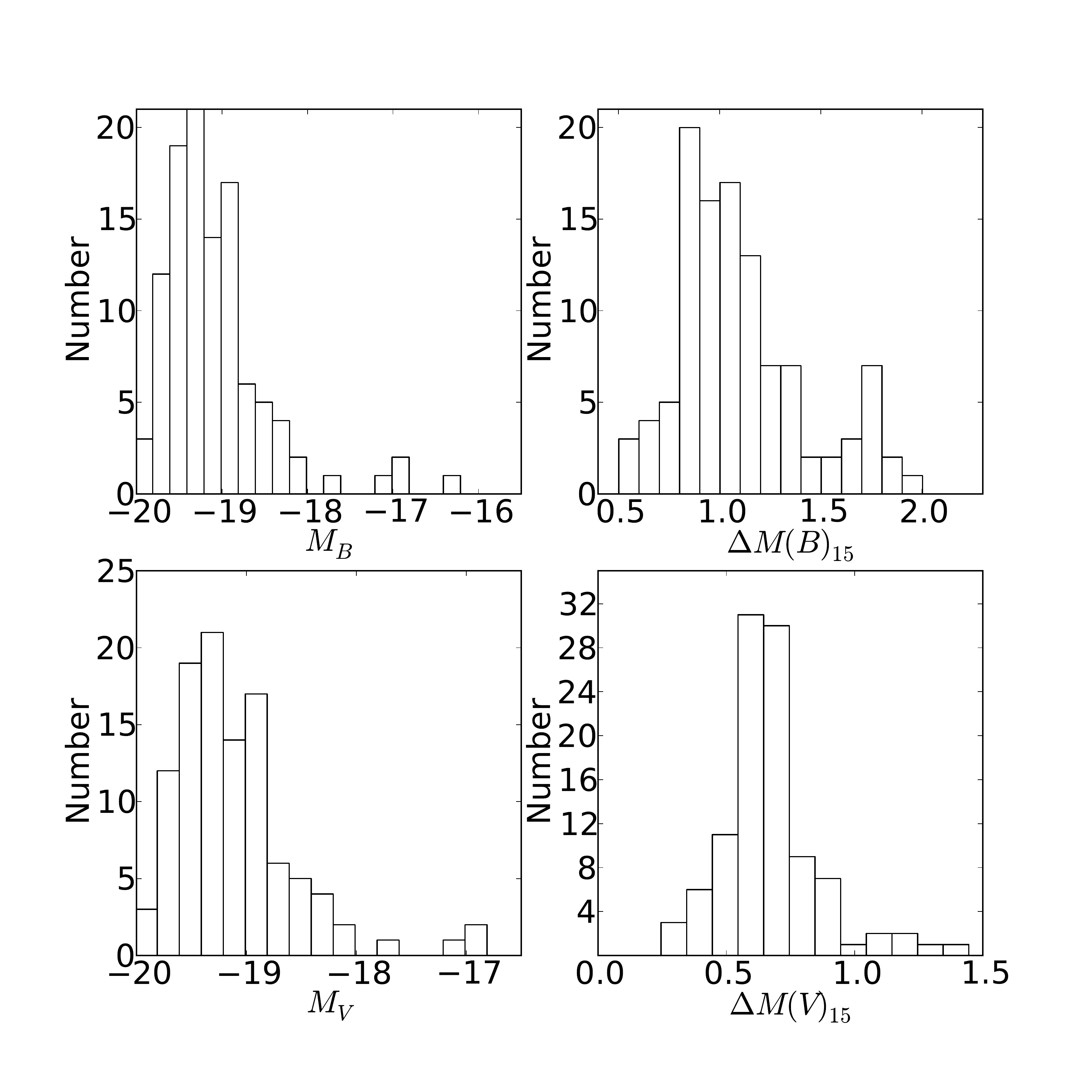}
\caption{\emph{Top:}  $B$-band luminosity distribution, corrected for host galaxy extinction. The bin sizes are 0.15 and 0.1 \,mag for the $M_{B}$ and $\Delta m_{15}(B)$ plots respectively. \emph{Bottom:} $V$-band luminosity distribution, corrected for host galaxy extinction. The bin sizes are 0.15 and 0.1 \,mag for the $M_{V}$ and $\Delta m_{15}(V)$ plots respectively.}
\label{fig:Bhostcorect}
\end{figure*}
\subsection{Host galaxy extinction}

Figure \ref{fig:hostextinction} shows the amount of host extinction for each
galaxy type derived with the method of \citet{Sasdelli16}. The top left plot
shows how the LD of SNe\,Ia from S-F galaxies changes when host galaxy
extinction is corrected for. The top-right plot shows that there is a large
difference between the values of $M_{B}$ before and after extinction 
correction. If there was no correction for extinction all data points would fall
on a linear relation. This is however not the case: 67\% of SNe\,Ia from S-F
galaxies are affected by detectable amounts of host galaxy extinction. The
average $E(B-V)$ of all SNe\,Ia from S-F host galaxies is $0.130\pm0.023$\,mag.
In contrast, only 43\% of the SNe\,Ia from passive galaxies are affected by
detectable amounts of host galaxy extinction, as would be expected due to the
lack of dust in early-type galaxies. Their mean $E(B-V)=0.040\pm0.013$\,mag.
Therefore on average SNe from passive galaxies have almost negligible host
galaxy extinction. The average change in absolute $B$ magnitude for SNe\,Ia from
S-F galaxies is $0.50$\,mag, compared to $0.15$\,mag for SNe from passive
galaxies.

\subsection{Luminosity Distribution}

The top left panel of Figure \ref{fig:Bhostcorect} shows the $M_B$ distribution
of the SNe after correction for host galaxy extinction. The distribution has 
mean $\bar{M_B}=-19.09\pm0.06$\,mag and a standard deviation
$\sigma(M_{B})=0.62$\,mag. The LD consists of a `normal' distribution of SNe\,Ia
and of a sub-luminous `tail' comprising faster declining SNe\,Ia. A similar
result is found in the $V$-band (Figure 7, lower left), where the distribution
has mean $\bar{M_V}=-19.10\pm$0.05\,mag and $\sigma(M_{V})=0.52$\,mag. From
Figure \ref{fig:Bhostcorect} it is apparent that when corrected for extinction
`normal' SNe\,Ia lie within the range $-20>M_{B}>-18$, and similarly for $M_V$. 

The right hand side of Figure \ref{fig:Bhostcorect} shows the distribution of
decline rates. In the $B$ band, the distribution has a mean $\Delta m_{15}(B) =
1.12\pm0.03$\,mag and a standard deviation of 0.32. A bimodal distribution is still
visible in $\Delta m_{15}(B)$  showing a lack of `transitional' objects. Table
4 contains the statistics of the SNe\,Ia sample when
corrected for extinction.

\subsection{LD by host galaxy type}

\begin{figure*}
\centering
\includegraphics[scale=0.7]{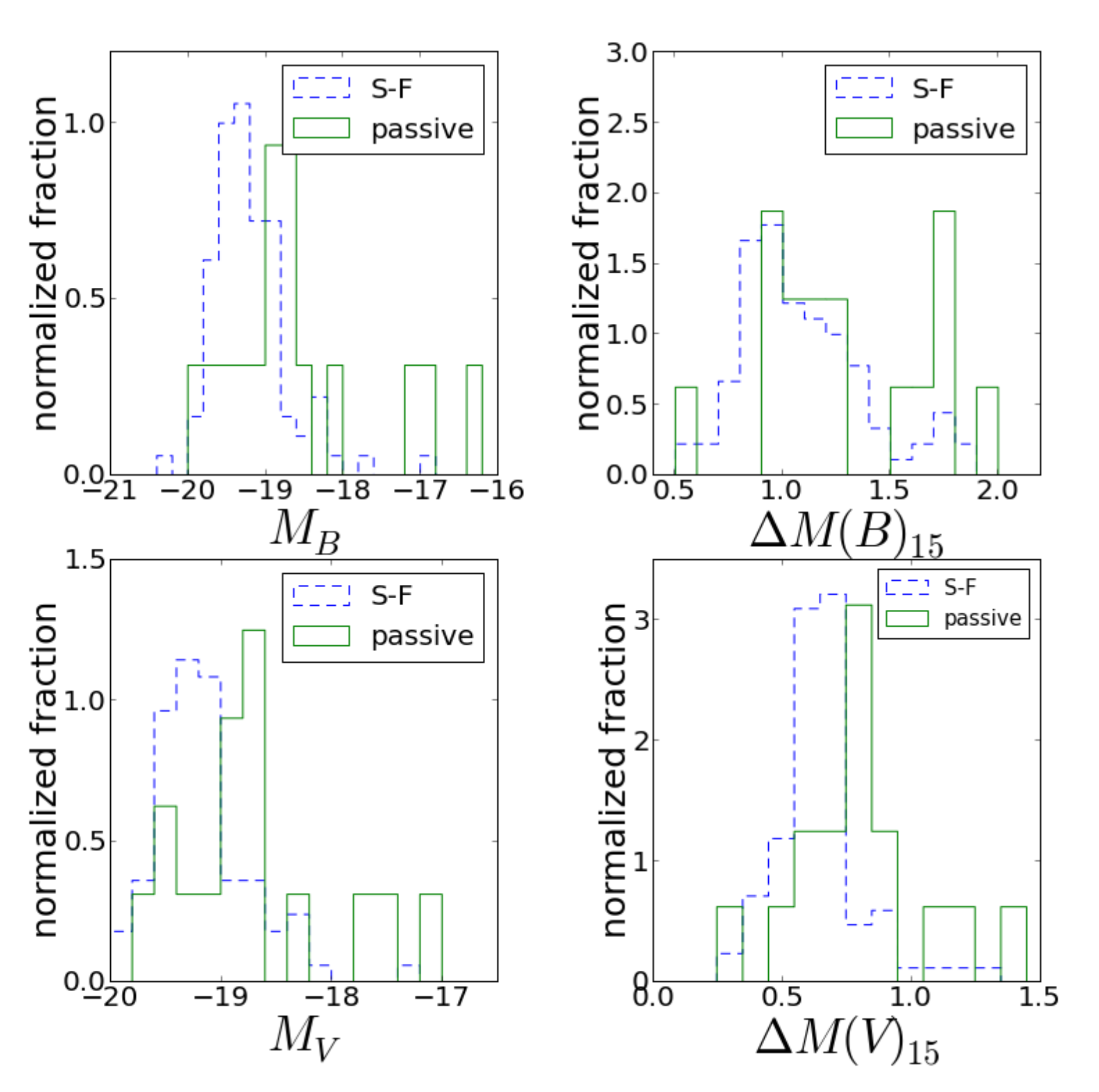}
\caption{SNe\,Ia separated by host galaxy type, after correction fot host galaxy extinction. Blue shows SNe\,Ia from S-F galaxies and green is SNe\,Ia from passive galaxies.}
\label{fig:Bhostcorectcore}
\end{figure*}

We separated the SNe by host galaxy type, as discussed in Section 2.  Figure
\ref{fig:Bhostcorectcore} shows the luminosity distributions of SNe\,Ia
separated by host galaxy type, after correction for extinction. In the $B$-band,
SNe from S-F galaxies have $\bar{M_B}=-19.20\pm0.05$\,mag, while SNe from
passive galaxies have $\bar{M_B}=-18.57\pm0.24$\,mag, which is
$0.63\pm0.24$\,mag dimmer.   In the $V$-band, SNe\,Ia from S-F galaxies have
$\bar{M_V}=-19.19\pm0.05$\,mag, while SNe from passive galaxies have
$\bar{M_V}=-18.71\pm0.18$\,mag.  The difference in $V$-band average absolute
magnitude is $0.48\pm0.24$\,mag.  Faster declining SNe tend to favour passive
galaxies, which is expected as sub-luminous SNe tend to favour old stellar
populations \citep{Howell01}. The mean $\Delta m_{15}(B)$ for SNe from passive
galaxies is $1.29\pm0.10$\,mag, compared to $1.10\pm0.03$\,mag for SNe\,Ia from
S-F galaxies. 

We ran K-S tests on the distributions from each host galaxy, see Table 5. The
probability that the $M_{B}$ samples come from the same parent distribution is
less than 0.1\%. There is marginal evidence for a difference between the
$\Delta m_{15}(B)$ distributions, with a 12\% probability that they came from
the same parent sample. When we ran these tests on the $V$-band, we found that
the $M_{V}$ distributions had less than a 0.4\% probability of coming from the
same parent distribution. 

The mean colour term for SNe\,Ia, after correction for host galaxy extinction is
$-0.008\pm0.013$\,mag. SNe\,Ia from S-F galaxies have an average
$(B-V)_{B\rmn{max}}=-0.025\pm0.010$\,mag. This is $0.12\pm0.06$\,mag bluer  than
for SNe\,Ia from passive galaxies, which have an average colour term of 
0.095$\pm$0.060\,mag.  This demonstrates that SNe\,Ia from S-F galaxies are
intrinsically bluer than those from passive galaxies.

\subsection{WLR after extinction correction}

\begin{figure}
\centering
\includegraphics[scale=0.35]{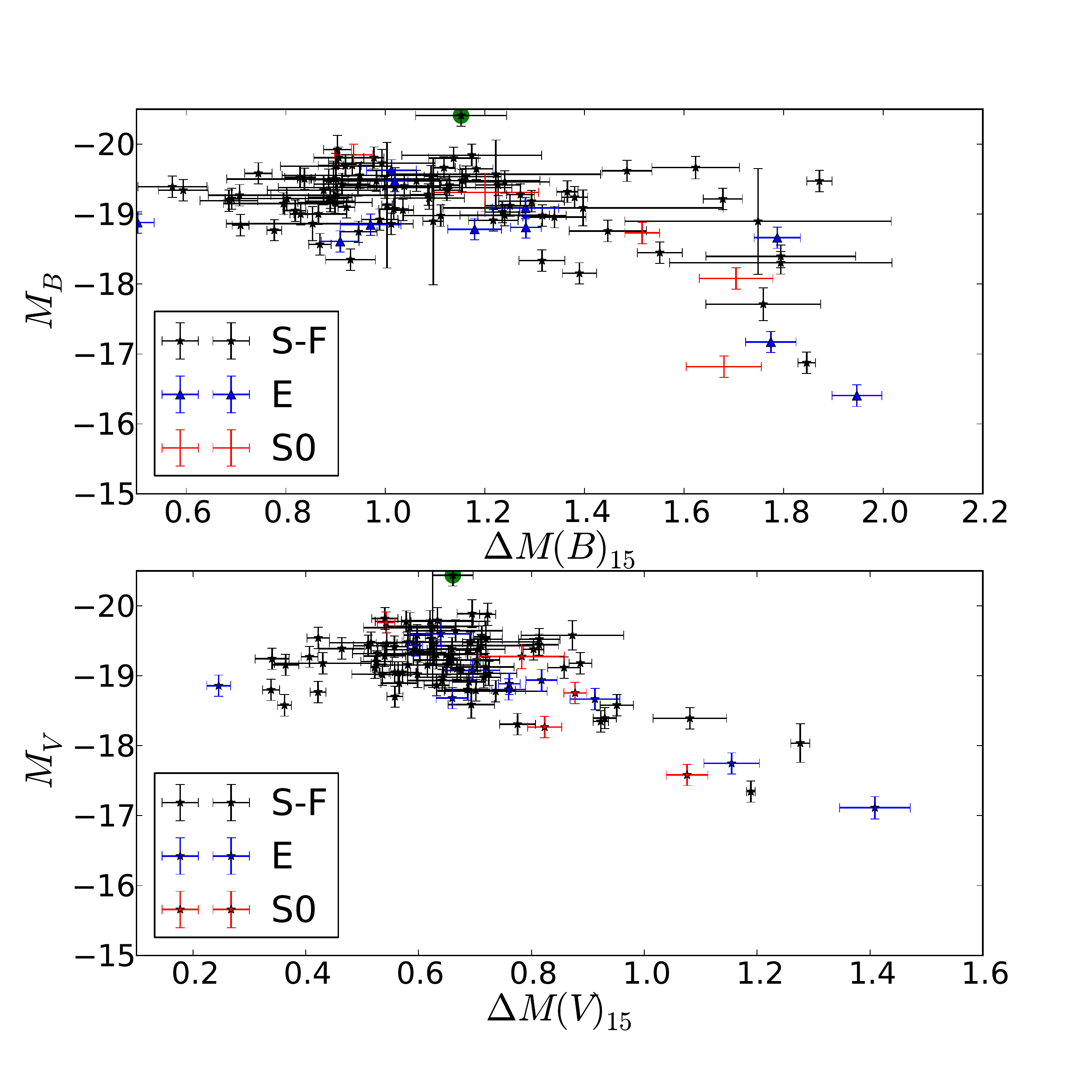}
\caption{The WLR corrected for host galaxy extinction. The black points are SNe 
from S-F galaxies, the blue are points are SNe from elliptical galaxies and the 
red points are SNe\,Ia from S0 galaxies. The green point is SN2003cg. 
\emph{Top:}  $B$-band \emph{Bottom:}  $V$-band}
\label{fig:BhostcorectWLR}
\end{figure}

Figure \ref{fig:BhostcorectWLR} shows the WLR of the SNe\,Ia in our sample after
correction for host galaxy extinction. There is a larger intrinsic scatter in
this WLR compared to when only SNe with $(B-V)_{B\rmn{max}}<0.01$\,mag are used.
This indicates that the parameter space that SNe\,Ia can fill is much larger
than originally thought. This is the result of the breaking of the
colour/extinction degeneracy by the method of \citet{Sasdelli16}. The bulk of
`normal' SNe\,Ia lie in a range of $\sim$1.5\,mag, rather than showing a tight
correlation between LC shape and absolute magnitude. Among SNe with $\Delta m_{15}(B)>$1.6\,mag, the sub-luminous tail
dominates. This is where transitional SNe, with 1.6$<\Delta m_{15}(B)<$1.8\,mag,
and subluminous SNe\,Ia, with $\Delta m_{15}(B)>$1.8\,mag, are. These SNe\,Ia
are far more diverse in properties than ``normal'' ones.  In Figure
\ref{fig:BhostcorectWLR} the highlighted green mark located well above the WLR
is SN\,2003cg. This is the SN with the largest extinction, $E(B-V)$=1.026\,mag.
\citet{EliasRosa06} find that SN\,2003cg has an anomalous extinction with 
$R_{v}=1.80\pm$0.19\,mag. SNe with very high extinction may require a different
reddening law, but this is the only case in our sample for which this occurs.
For more information on highly extinguished objects see \citet{Sasdelli16}.

 \begin{table*}
  \centering
 \begin{minipage}{120mm}
 \caption{Statistics of data before correction for host galaxy extinction. The data used in this table are a subset of the full sample.}
 \begin{tabular}{cccccc} 

 \hline 
   & All & S-F &  passive & E & S0 \\ 
Amount of SNe &109 &89 &16 &11 &5  \\  
 \hline 
$ \overline{M_{B}} $&--18.66 $\pm$0.07 & --18.70 $\pm$0.08 & --18.42 $\pm$0.23 & --18.48 $\pm$0.28 & --18.29 $\pm$0.43 \\  
 $\sigma{M_{B}}$ &0.74 & 0.71 & 0.94 & 0.92 & 0.97 \\  
 median$M_{B}$ &--18.85 $\pm$0.09 & --18.88 $\pm$0.09 & --18.80 $\pm$0.29 & --18.81 $\pm$0.35 & --18.71$\pm$0.54  \\  
$ \overline{\Delta m_{15}B}$ &1.11 $\pm$0.03 &1.08 $\pm$0.03 &1.29 $\pm$0.10 &1.24 $\pm$0.13 &1.40$\pm$0.13  \\  
$ \sigma{\Delta m_{15}B}$ &0.32 &0.29 &0.39 &0.42 &0.30  \\  
  \hline  
$\overline{(B-V)}$ &0.097 $\pm$0.022 &0.095 $\pm$0.025 &0.132 $\pm$0.062 &0.104 $\pm$0.070 &0.193 $\pm$0.118  \\  
 $\sigma{(B-V)}$ &0.233 &0.233 &0.247 &0.234 &0.263  \\  
 $median{(B-V)}$ &0.040 $\pm$0.03 &0.040 $\pm$0.03 &0.039 $\pm$0.08 &0.030 $\pm$0.09 &0.077 $\pm$0.15 \\  
  \hline  
 $\overline{M_{V}}$ &--18.77 $\pm$0.05 & --18.81 $\pm$0.06 & --18.60 $\pm$0.17 & --18.63 $\pm$0.20 & --18.53$\pm$0.31 \\  
 $\sigma{M_{V}}$ &0.55 & 0.52 & 0.68 & 0.68 & 0.68 \\  
 median $M_{V}$ &--18.88 $\pm$0.07 &--18.90 $\pm$0.07 &--18.83 $\pm$0.21 &--18.86 $\pm$0.26 &--18.74$\pm$0.38 \\   
$ \overline{\Delta m_{15}V}$ &0.67 $\pm$0.02 &0.65 $\pm$0.02 &0.80 $\pm$0.06 &0.79 $\pm$0.09 &0.82$\pm$0.08  \\  
$ \sigma{\Delta m_{15}V}$ &0.19 &0.16 &0.26 &0.29 &0.17  \\  
  \hline 
\end{tabular}

\end{minipage}
\label{table:Absmaghostcorbefore}
\end{table*}

  \begin{table*}
  \centering
 \begin{minipage}{120mm}
 \caption{Statistics of the data after correction for host galaxy extinction.}
 \begin{tabular}{cccccc} 
 \hline 
   & All & S-F &  Passive & E & S0 \\ 
Amount of SNe &109 &89 &16 &11 &5  \\  
 \hline 
$ \overline{M_{B}} $&--19.09 $\pm$0.06 & --19.20 $\pm$0.05 & --18.57 $\pm$0.24 & --18.58 $\pm$0.27 & --18.56 $\pm$0.47 \\  
$ \sigma{M_{B}} $&0.62 & 0.49 & 0.96 & 0.91 & 1.05 \\  
 median${M_{B}}$ &--19.22 $\pm$0.07 & --19.27 $\pm$0.07 & --18.80 $\pm$0.30 & --18.81 $\pm$0.34 & --18.73$\pm$0.59  \\  
 $\overline{\Delta m_{15}B}$ &1.12 $\pm$0.03 &1.10 $\pm$0.03 &1.29 $\pm$0.10 &1.24 $\pm$0.13 &1.41 $\pm$0.13  \\  
$ \sigma{\Delta m_{15}B}$ &0.32 &0.29 &0.39 &0.42 &0.30  \\  
  \hline  
  $\overline{(B-V)}$ &--0.008 $\pm$0.013 &--0.025 $\pm$0.010 &0.095 $\pm$0.060 &0.079 $\pm$0.067 &0.129 $\pm$0.119  \\  
 $\overline{(B-V)}$ &0.135 &0.097 &0.239 &0.223 &0.266  \\  
 $median(B-V)$ &--0.029 $\pm$0.02 & --0.034 $\pm$0.01 & --0.007 $\pm$0.07 & --0.019 $\pm$0.08 & 0.004$\pm$0.15  \\  
  \hline  
$ \overline{M_{V}} $&--19.10 $\pm$0.05 & --19.19 $\pm$0.05 & --18.71 $\pm$0.18 & --18.71 $\pm$0.20 & --18.73 $\pm$0.34 \\  
$ \sigma{M_{V}} $&0.52 & 0.45 & 0.70 & 0.68 & 0.76 \\  
 median $M_{V}$ &--19.17 $\pm$0.06 & --19.23 $\pm$0.06 & --18.83 $\pm$0.22 & --18.86$\pm$0.26 & --18.75$\pm$0.43  \\  
$ \overline{\Delta m_{15}V}$ &0.67 $\pm$0.02 &0.65 $\pm$0.02 &0.80 $\pm$0.07 &0.79 $\pm$0.087 &0.82$\pm$0.08  \\  
$ \sigma{\Delta m_{15}V} $&0.19 &0.16 &0.26 &0.29 &0.17  \\  
  \hline  
 $\overline{E(B-V)}$ &0.114 $\pm$0.019 &0.130 $\pm$0.023 &0.040 $\pm$0.013 &0.026 $\pm$0.013 &0.070 $\pm$0.026  \\  
 $\sigma{E(B-V)}$ &0.20 &0.21 &0.05 &0.04 &0.06  \\  
  \hline 
\end{tabular}
\end{minipage}
\label{table:Absmaghostcor}
\end{table*}

 \begin{table}
  \centering
 \caption{ K-S tests for SNe with different types of host galaxies, after correction for host galaxy extinction.}
 \begin{tabular}{ccc} 
 \hline 
 Compared distributions & Value used & $P$ value \\
 \hline 
SNe active vs SNe passive& $M(B)_{15}$& 0.001 \\
SNe active vs SNe passive& $\Delta M(B)_{15}$& 0.118 \\
SNe active vs SNe passive& $M_{V}$ & 0.004 \\
 \hline 
\end{tabular}
\label{table:KShostdo}
\end{table}

\section{Full luminosity distribution}

\begin{figure*}
\centering
\includegraphics[scale=0.3]{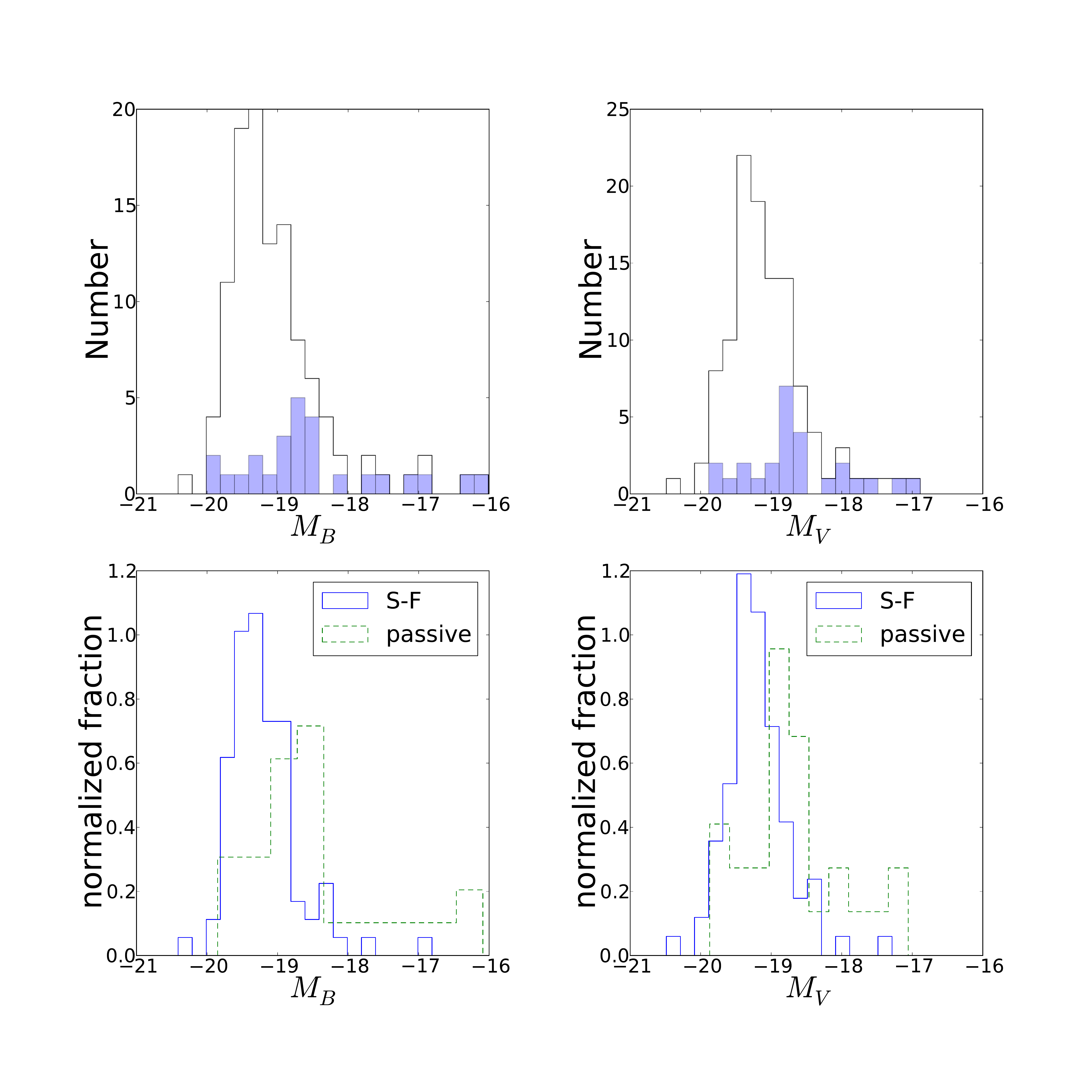}
\caption{The above plots are the final $M_{B}$ and $M_{V}$ luminosity distribution plots for the sample used in this paper. SNe\,Ia from passive galaxies with no known extinction have been re-added to this distribution to increase the sample size.   \emph{Top:} The left plot is the full $M_{B}$ luminosity distribution, the right plot is the full $M_{V}$ luminosity distribution. The overlaid blue histograms are the distributions of SNe from passive galaxies.  \emph{Bottom:} The left plot is the full $M_{B}$ luminosity distribution separated by host galaxy type, and the right plot is the full $M_{V}$ distribution separated by host galaxy type.}
\label{fig:fullLDplot}
\end{figure*}

\begin{figure*}
\centering
\includegraphics[scale=0.3]{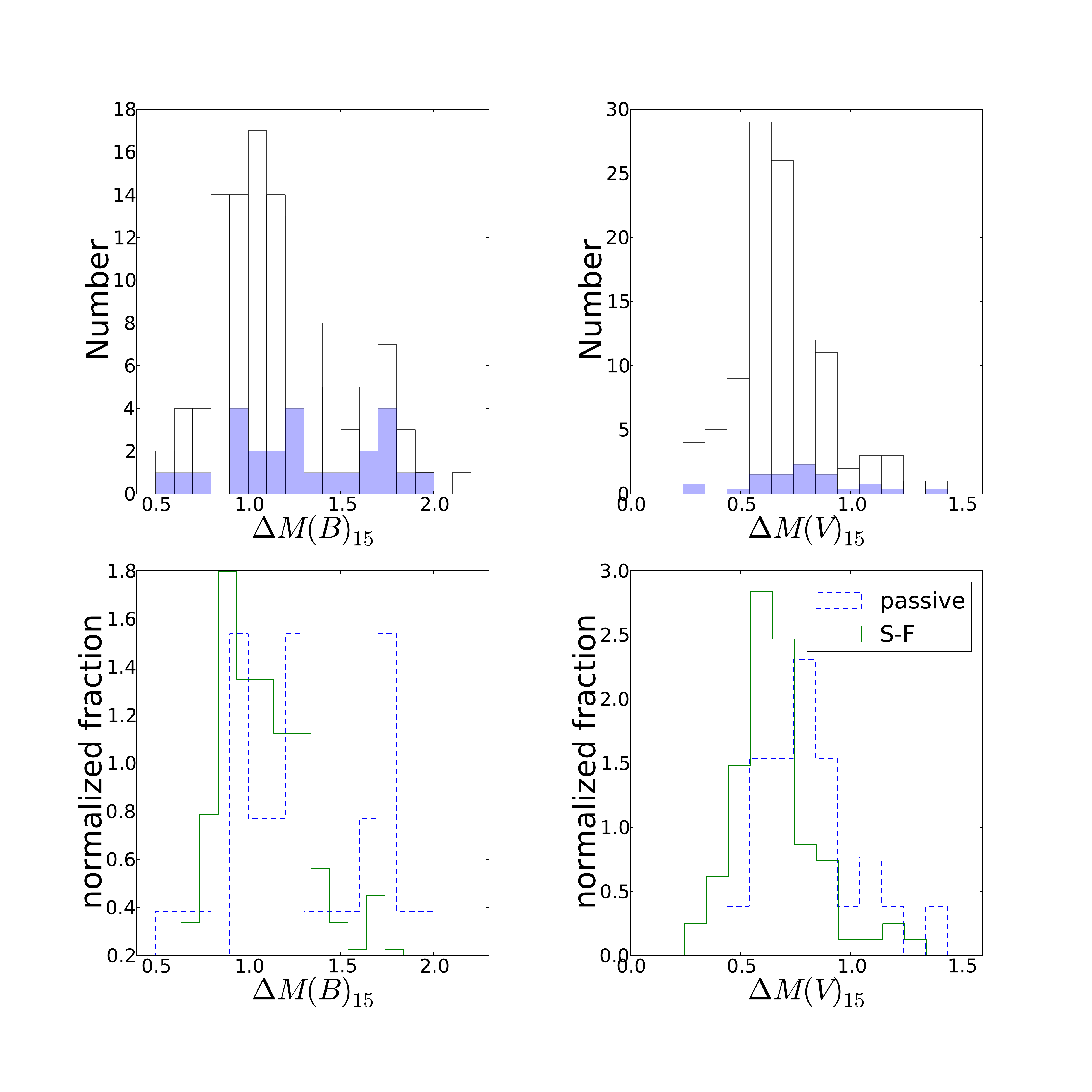}
\caption{The final $\Delta M_{15}(B)$ and $\Delta M_{15}(V)$ LD plots for the sample used in this paper. SNe\,Ia from passive galaxies with no known extinction have been re-added to this distribution to increase the sample size.\emph{Top:} The left plot is the full $\Delta m_{15}(B)$ LD, the right plot is the full $\Delta M_{15}(V)$ LD. The overlaid blue histograms are the distributions of SNe from passive galaxies. \emph{Bottom:} The left plot is the full $\Delta M_{15}(B)$ LD separated by host galaxy type, and the right plot is the full $\Delta m_{15}(V)$ distribution separated by host galaxy type.}
\label{fig:fullLDplotdm15}
\end{figure*}

It is apparent that SNe\,Ia from passive galaxies suffer only from negligibly
from host galaxy extinction, with 57\% of SNe\,Ia from passive galaxies having
no detectable host galaxy extinction. The mean $E(B-V)$ for SNe\,Ia from passive
galaxies is $0.04\pm0.013$\,mag. In this section we use the sample of 105
SNe\,Ia which could be corrected for host galaxy extinction, and add back in the
10 SNe\,Ia from passive galaxies for which host galaxy extinction could not be
derived. We do this to avoid small sample statistics. We justify this step
because it is reasonable to assume that they SNe in passive galaxies have very
little host extinction\footnote{The one exception is SN 1986G which was
corrected for extinction of $E(B-V)$=0.65\,mag \citep{Nugent05}. SNe 1986G is from an S0 galaxy,
but sits in the dust lane in NGC 5128.}. The new sample sizes comprises then 115
SNe, of which 89 are from S-F galaxies and 26 from passive galaxies. Figure
\ref{fig:fullLDplot} is the final SN\,Ia luminosity distribution. The mean
absolute magnitudes of the full sample are $\bar{M_B}=-19.04\pm0.07$\,mag and
$\bar{M_V}=-19.07\pm0.06$\,mag. Figure \ref{fig:fullLDplotdm15} shows the final
SNe\,Ia $\Delta m_{15}$ distributions. 

In Figure \ref{fig:fullLDplot} we plot the LD of SNe\,Ia, divided by host galaxy
type. We can assume that at least three different populations of SNe\,Ia are
present in these plots: a normal population of SNe\,Ia from S-F galaxies, which
dominates the population of normal SNe, a normal population of SNe\,Ia from
passive galaxies, and a subluminous population, which is dominated by SNe in
passive galaxies. The numbers of subluminous SNe in S-F galaxies is too small to
determine whether their population has different properties with respect to
that in passive galaxies.

 \begin{table}
  \centering
 \caption{Statistics of full LD.}
 \begin{minipage}{120mm}
 \begin{tabular}{cccc} 
 \hline 
 & All & S-F &  passive \\
 Amount of SNe &115 &89 &26  \\ 
 \hline 
 $ \overline{M_{B}}$ &--19.04 $\pm$0.07 &--19.20 $\pm$0.05 &--18.48 $\pm$0.19 \\ 
 $ \sigma{M_{B}}$ &0.70 &0.49 &0.98 \\
 median${M_{B}}$ &--19.20 $\pm$0.08 &--19.27 $\pm$0.07 &--18.68 $\pm$0.24\\ 
 \hline 
 $ \overline{M_{V}}$ &--19.07 $\pm$0.06 &--19.19 $\pm$0.05 &--18.67 $\pm$0.14 \\ 
 $ \sigma{M_{V}}$ &0.57 &0.45 &0.72 \\ 
 median${M_{V}}$ &--19.16 $\pm$0.07 &--19.24 $\pm$0.06 &--18.78 $\pm$0.18 \\ 
 \hline 
 $ \overline{M_{B}}$ \footnote{$M_{B}$<-18\,mag} &--19.19 $\pm$0.04 &--19.24 $\pm$0.04 &--18.94 $\pm$0.11 \\ 
 $ \sigma{M_{B}}$ &0.43 &0.40 &0.47 \\
 median${M_{B}}$ &--19.23 $\pm$0.05 &--19.27 $\pm$0.05 &--18.80 $\pm$0.13\\ 
 \hline 
 $\overline{M_{V}}$ \footnote{$M_{V}$<-18\,mag} &--19.16 $\pm$0.04 &--19.22 $\pm$0.04 &--18.94 $\pm$0.10 \\ 
 $ \sigma{M_{V}}$ &0.43 &0.41 &0.46 \\ 
 median${M_{V}}$ &--19.19 $\pm$0.05 &--19.25 $\pm$0.06 &--18.86 $\pm$0.12 \\ 
 \hline 
\end{tabular}
\end{minipage}
\label{table:Absmaghostcorfull}
\end{table}

The shape of the LD depends strongly on host type. 
The LDs of SNe\,Ia in S-F galaxies have Gaussian shapes, with mean magnitudes 
$\bar{M_B}=-19.20\pm$0.05\,mag and $\bar{M_V}=-19.19\pm0.05$\,mag and 
standard deviations of 0.49 and 0.45\,mag, respectively. 
In passive galaxies the SN\,Ia LD is much wider: the mean magnitudes are
$\bar{M_B}=-18.48\pm0.19$=\,mag and $\bar{M_V}=-18.67\pm0.14$\,mag, with   
standard deviations 0.49\,mag and 0.98\,mag respectively. 

To compare the distributions of `normal' SNe\,Ia from the two different host
galaxy types we must exclude subluminous SNe\,Ia.  This is done by removing any
SNe\,Ia with $M_B, M_V > -18.0$\,mag. These values are obtained by visual
inspection of the plots. It should be noted that the cut off is applied in terms
of absolute magnitude rathen than decline rate, as fast declining SNe\,Ia can be
both intrinsically bright and intrinsically dim (Figure
\ref{fig:BhostcorectWLR}). The results are shown in Figure \ref{fig:fullLDplot}.

When the subluminous population is removed, the distributions change.  S-F
galaxies have $\bar{M_B}=-19.24\pm$0.04\,mag and 
$\bar{M_V}=-19.22\pm$0.05\,mag, with median values $M_B=-19.27\pm$0.05\,mag and 
$M_V=-19.25\pm$0.06\,mag.   For passive galaxies,
$\bar{M_B}=-18.94\pm0.11$\,mag,  $\bar{M_V}=-18.94\pm$0.10\,mag, with median
values $M_B=-18.80\pm0.13$\,mag and $M_V=-18.86\pm0.12$\,mag. 

The photometric distributions of `normal' SNe\,Ia depends therefore on host
galaxy types.  The differences between the mean values of $M_{B}$ and $M_{V}$
are 0.30$\pm$0.11\,mag and  0.28$\pm$0.11\,mag respectively, and are
0.47$\pm$0.14\,mag  and  0.39$\pm$0.13\,mag in the median values. These two
separate distributions lead us to the conclusion that within the  `normal'
SNe\,Ia population, intrinsically dimmer SNe\,Ia are more common from older
stellar populations, confirming previous results {\citep[e.g.][]{Sullivan06}}.

\section{SNe\,Ia from young and old stellar populations}

SNe\,Ia are thought to come from both young ($<400 Myr$) and old ($>2.4Gyr$)
stellar populations \citep{Brandt10}. Traditionally, faster declining SNe\,Ia
are thought to come from old stellar populations, and slower declining SNe\,Ia
from young stellar systems. In section 5 we have determined that there are at
least 3 populations of SNe\,Ia, `normal' SNe from S-F, `normal' SNe from passive
galaxies and subluminous SNe. As passive galaxies predominantly consist of old
stars, we conclude that some of the `normal' SNe\,Ia must originate from old
stellar systems. On the other hand, S-F galaxies consist of both young and old
stellar populations. Therefore they must contain SNe from both groups. In this
section we attempt to quantify the fraction of SNe\,Ia from S-F galaxies which
are produced from old stellar populations. From this the LD of SNe\,Ia from
young stellar populations is produced.     

\begin{figure*}
\centering
\includegraphics[scale=0.3]{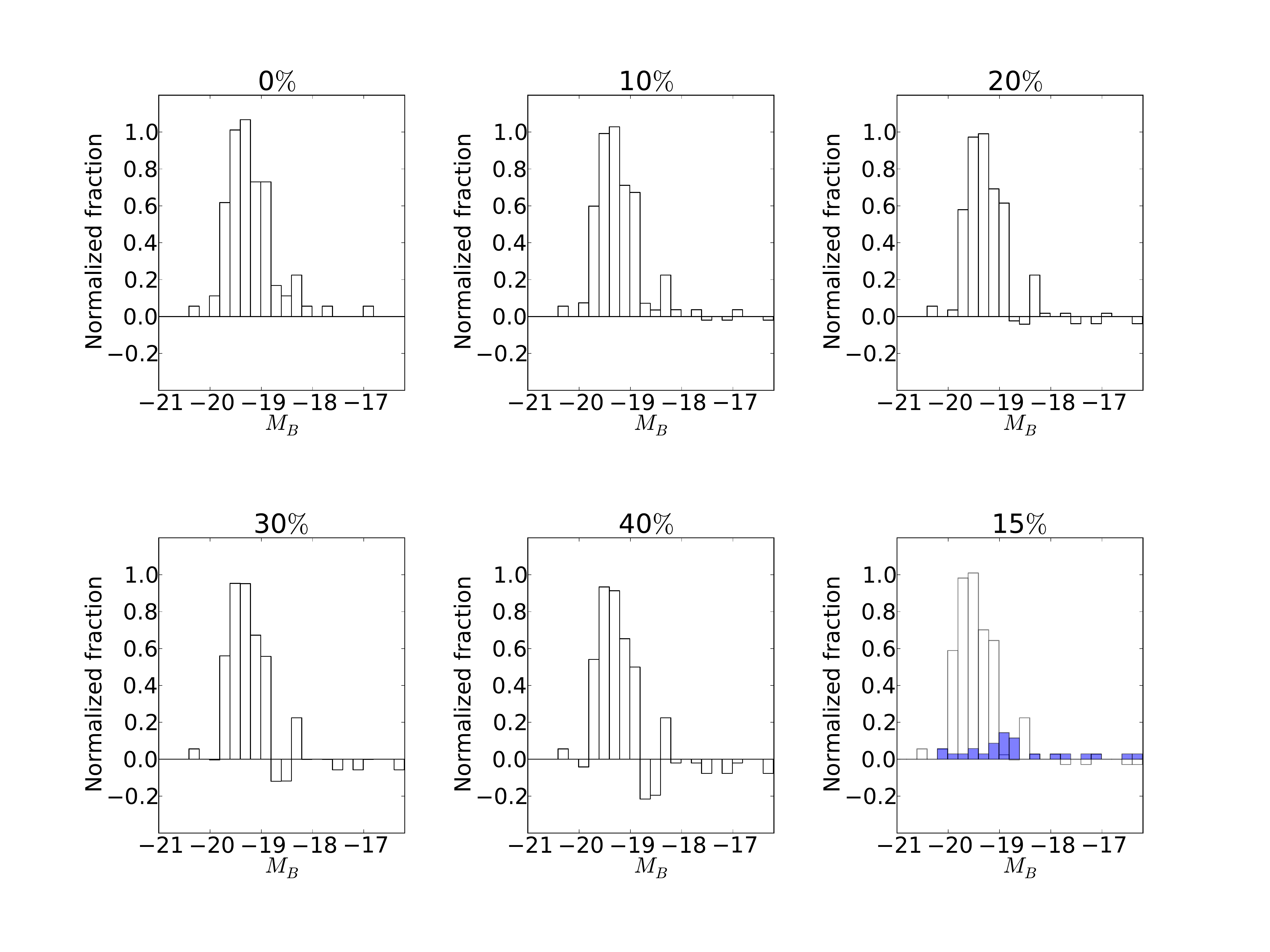}
\caption{The $B$-band LD for SNe\,Ia from passive galaxies normalized, scaled and subtracted from the LD of SNe\,Ia from S-F galaxies. The factor by which the SNe\,Ia from passive galaxies has been normalized is presented on the top of each plot. The blue bars in the bottom right panel are the LD of SNe\,Ia from passive galaxies scaled to 15\% of those from active galaxies.}
\label{fig:fullbsubhist}
\end{figure*}

\begin{figure*}
\centering
\includegraphics[scale=0.3]{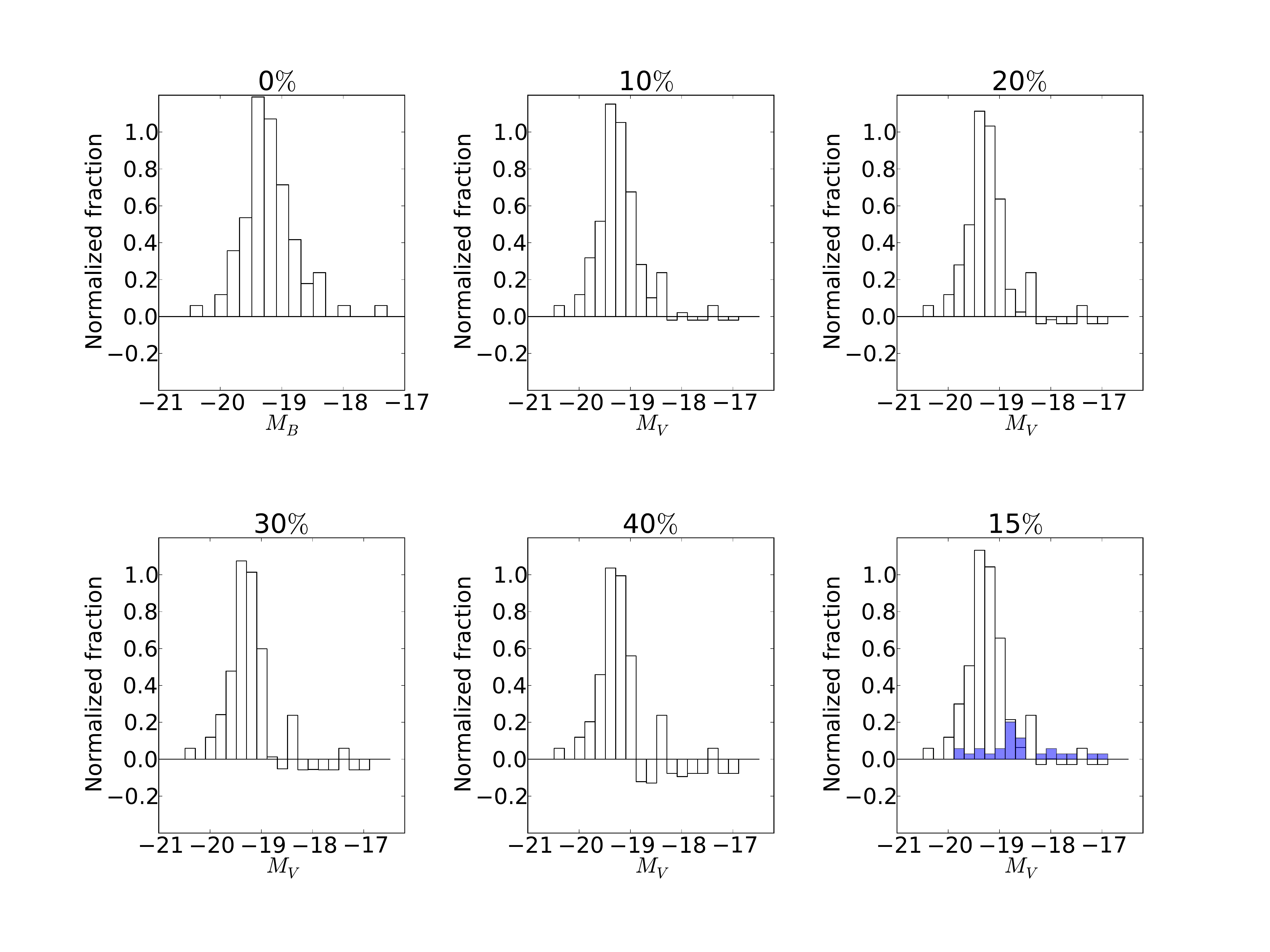}
\caption{The $V$-band LD for SNe\,Ia from passive galaxies normalized, scaled and subtracted from the LD of SNe\,Ia from S-F galaxies. The factor by which the SNe\,Ia from passive galaxies has been normalized is presented on the top of each plot. The blue bars in the bottom right panel are the LD of SNe\,Ia from passive galaxies scaled to 15\% of those from active galaxies.}
\label{fig:fullvsubhist}
\end{figure*}

We use the normalised LDs shown in Figure \ref{fig:fullLDplot} to determine the
fraction of SNe\,Ia from young/old stellar populations. This is done by scaling
the LD from passive galaxies and subtracting this distribution from SNe from S-F
galaxies. It is known that SNe\,Ia with $M_{B}>-18$\,mag are mostly from older
stellar populations, so we examine only SNe\,Ia with $M_{B}<-18$\,mag. Both LDs
were divided into 20 bins with similar size (0.2\,mag); so that the LDs can be 
scaled and subtract from one another. We scaled the passive LD to 10\%, 20\%,
30\% and 40\% with respect to the S-F LD and subtract it. We are assuming that
our sample of passive SNe\,Ia is a fair representation of the intrinsic sample,
i.e. that SNe from passive environments in S-F galaxies have the same properties
as SNe from passive galaxies. The scaled and subtracted S-F LDs are shown in
Figure \ref{fig:fullbsubhist}, which shows that if the passive-like component
exceeds 20 per cent of the total the subtraction is unrealistic as there are too
many negative bins. Therefore, in the $B$-band we select a value of
$(15\pm10)\%$ as the optimum scaling value to use in the LD subtractions, and
suggest that this is the fraction of `normal' SNe\,Ia in S-F galaxies which are
from old stellar populations. The LD when a scaling factor of 15\% is used is
shown in the bottom right panel in Figure \ref{fig:fullbsubhist}, where we also
show the ``passive'' component.  In the remaining plots of Figure
\ref{fig:fullbsubhist}, we only show the result of the subtraction, i.e. the
``young'' component. The LD of SNe\,Ia from young stellar systems is narrower
than that of he whole sample.  This effect is seen in both the $B$ and $V$
bands. Figure \ref{fig:fullvsubhist} shows the $V$-band scaled and subtracted
LDs.  The $V$-band results also suggest that $(15\pm10)\%$ of `normal' SNe\,Ia
from S-F galaxies are from old stellar populations.  The bottom right panel in
Figure \ref{fig:fullvsubhist} presents the $V$-band LD when a scaling factor of
15\% is chosen. The blue bars overlaid in this plot show the LD of SNe\,Ia from
passive galaxies scaled to 15\%.  Therefore we conclude that there are
photometrically 4 populations of SNe\,Ia: `normal' ones from S-F galaxies that
come from young stellar systems, `normal' SNe\,Ia from passive galaxies, which
come from an old population, `normal' SNe\,Ia from S-F galaxies that come from
an old population, and subluminous SNe, which are thought to come from old
stellar systems.

If we assume that our sample is complete, and use  $M_{B}<-18$\,mag as the cut off between 
subluminous and normal SNe Ia, the frequency of SNe Ia in our sample can be quantified.
Out of the 115 SNe Ia in Figure \ref{fig:fullLDplot}, 8 (7\%) are subluminous. This leaves 107 SNe Ia, of which 20 
are  `normal' SNe\,Ia from passive galaxies (from an old population), and 16  are `normal' SNe\,Ia from S-F galaxies and come from
an old population. Therefore, 32 (28\%) of all SNe Ia are normal and come from an old population.  Which means, 75 (65\%) of SNe Ia are normal 
and come form a young stellar population.

\section{Discussion}

The aim of this work was to explore the diversity of SN\,Ia the LCs. Therefore,
we built luminosity distributions using as few prior assumptions as possible. We
have not relied on any empirical LC fitting method, and used distances to the
SNe derived from independent methods.  We chose to use mean values of $\Delta
M_{15}(B/V)$ and $M_{B/V}$ for our analysis because this is a good way to
highlight outliers, which make up a small fraction of the overall population.
The analysis is presented both before and after applying a corrected for host
galaxy extinction. Deriving the host galaxy extinction of SNe\,Ia is a difficult
task as it requires resolving the degeneracy between colour and extinction. 
Although the method used for calculating the host galaxy extinction can resolve
the colour-extinction degeneracy, it does remove the intrinsically unusual SNe,
as there are not enough unreddened spectral matches. Only a subset of the SNe
can be corrected for extinction, and some peculiar SNe may not pass this cut.
Therefore, the true diversity of SNe\,Ia may be underrepresented in the
corrected sample. 

Before correction for host galaxy extinction, our sample includes 165 SNe\,Ia,
134 of which are from S-F galaxies, 26 from passive galaxies and 5 from host
galaxies whose type could not be determined. The average $M_{B}$ and $M_{V}$ for
the sample are $-18.58\pm$0.07\,mag and $-18.72\pm$0.05\,mag, respectively. The
average $\Delta m_{15}(B)$ and $\Delta m_{15}(V)$ values for the sample are
1.14$\pm$0.03\,mag and 0.68$\pm$0.01\,mag. 

We find a bimodal distribution in $\Delta m_{15}(B)$ (Figure \ref{fig:LDB}),
with a lack of transitional SNe. Transitional SNe such as 1986G, 2003hv and
iPTF13ebh all have unique properties, and can hold information about potential
differences between progenitor channels of sub-luminous and normal SNe\,Ia. If
it is not a selection effect, the lack of transitional SNe could suggest that 
SNe\,Ia may come from multiple populations, with different LC properties, which
only cross in $\Delta m_{15}(B)$ at their extremes.  

Correcting the data for host galaxy extinction reduces the sample size to 109
SNe\,Ia, 89 of which are from S-F galaxies, 16 from passive galaxies and 5 from
host galaxies whose type could not be determined.  The smaller sample is
comparable to the larger sample in most relevant parameters.  SNe\,Ia from S-F
galaxies suffer from much larger host extinction than those from passive
galaxies: 67\% of the SNe\,Ia from S-F galaxies are affected by detectable host
galaxy extinction, whose the average value is $E(B-V)$=0.130$\pm$0.023\,mag. In
contrast, only 43\% of the SNe\,Ia from passive galaxies are affected by
detectable extinction, with average $E(B-V)=0.04\pm$0.013\,mag.  When corrected
for host galaxy extinction, the mean $\bar{M_B}$=-19.09$\pm$0.06\,mag and
$\bar{M_V}$=-19.10$\pm$0.05\,mag. In the $B$-band SNe from passive galaxies were
found to be 0.63$\pm$0.25\,mag less luminous on average than SNe in S-F
galaxies. The same is true in the $V$-band, where the difference is
0.48$\pm$0.19\,mag. 

The WLR is one of the underlying foundations in SN\,Ia cosmology. It is driven
by Ni mass, opacity and ejecta mass. We find a strong correlation between 
$\Delta m_{15}(B)$ and absolute magnitude for normal SNe when only SNe with have
a small observed colour term are selected. The WLR in our study is similar to
those of \citet{Phillips93} and \citet{phillips99}. However, after correcting
for host galaxy extinction the spread of properties increases, as shown in
Figure \ref{fig:BhostcorectWLR}. Furthermore, correcting for reddening leads to
the removal of some of the more unusual SNe\,Ia, and so the true parameter space
SNe\,Ia can fill in the WLR diagram is probably even larger.   

K-S tests were run on the sample after correction for extinction. The
probability that $M_{B}$ values for SNe\,Ia from passive and S-F galaxies come
from the same parent distribution was found to be $< 0.1$\%. In the case of
$M_{V}$ the probability is $< 0.4$\%. The likelihood that the $\Delta m_{15}(B)$
samples come from the same parent distributions is $\simeq$12\%.  

SNe\,Ia in passive galaxies for which the extinction could not be derived were
reintroduced to the extinction-corrected sample. As only 43\% of the SNe\,Ia
from passive galaxies were affected by detectable host galaxy extinction, and
this average extinction was very small, we assumed that the reintroduced sample
has negligible host extinction.  When the final LD is separated by host galaxy
type, it is found that there are three main populations of SNe\,Ia: `normal'
SNe\,Ia in S-F galaxies, `normal' SNe\,Ia in passive galaxies and sub-luminous
SNe\,Ia. 
Normal SNe\,Ia in S-F galaxies were found to have a median 
$M_{B}$=--19.27$\pm$0.05\,mag and $M_{V}$=--19.25$\pm$0.06\,mag.
Normal ($M_{B}<-18$) SNe\,Ia in passive galaxies have median
$M_{B}$=--18.80$\pm$0.13\,mag and $M_{V}$=--18.86$\pm$0.12\,mag.  
The difference hints at differences in properties of normal SNe\,Ia depending on 
host galaxy type, as shown in Figure \ref{fig:fullLDplot}.
SNe\,Ia in S-F galaxies appear to be a much more uniform sample: their 
distribution is similar to a Gaussian, whereas SNe\,Ia in passive galaxies 
appear to be far more diverse. 

In the final section of the analysis we normalise the LDs for SNe\,Ia from
passive and S-F galaxies. In order to quantify the fraction of `normal' SNe\,Ia
which come from old stellar systems in S-F galaxies we scaled the LD of SNe\,Ia
from passive galaxies to represent different fractions of the full S-F LD and
subtracted it from the full S-F LD. We find that $(15\pm10)\%$ of `normal'
SNe\,Ia from S-F galaxies are likely to come from old stellar systems.

In order fully to explore the range in parameter space that SNe\,Ia can fill one
would need to obtain a large sample LF, in both redshift and volume, which is out of the scope of our study.
\citet{Yasuda10} produced a LF, however they removed any peculiar SNe\,Ia and did not break the
colour-reddening degeneracy. \citet{Li2011} produced a volume-limited LF, but
did not correct for host galaxy extinction. 

As transitional objects appear to be rare, a large sample survey would have to
operate continuously for many years to produce a comprehensive LF of SNe\,Ia. In
the years of LSST and large high cadence surveys, along with robotic telescopes
on multiple sites (LCOGT), this will be possible. We will then be able to know
the variance of SNe\,Ia and place tighter constraints on the progenitor
systems.

\section{Conclusions}

This paper takes a different approach to SN\,Ia LC analysis. We made as few
assumptions as possible about the SNe, so two SNe\,Ia with a similar LC shape
can have different properties. This way it is possible to explore the diversity
of SNe\,Ia. We studied SNe\,Ia by first independently obtaining the distance to
each one then calculating its properties, $M_{B}$, $\Delta m_{15}(B)$, $M_{V}$
and $\Delta m_{15}(V)$. This allowed us to see the parameter space that SNe\,Ia
can fill.

We confirm previous results that SNe in passive galaxies tend to declinie more
rapidly, with an average $\Delta m_{15}(B)=1.29\pm0.08$\,mag, compared to
$\Delta m_{15}(B)= 1.11\pm0.03$\,mag for SNe in S-F galaxies. This is partly due
to the presence of sub-luminous SNe\,Ia, which are mostly found in older stellar
populations. Our method confirms that SNe with small observed 
$(B-V)_{B\rmn{max}}<0.01$\,mag follow the WLR, with some scatter. However, when
a larger sample of SNe are corrected for host galaxy extinction the scatter in
the WLR increases. We find that the range in $M_{B}$ of normal SNe\,Ia in the
WLR is $\sim$1.5\,mag. This suggests that (intrinsically) SNe\,Ia can fill a large
parameter space on the WLR. The distribution in $\Delta m_{15}(B)$ was found to
be bimodal, with a lack of transitional SNe. More data from transitional SNe can
hold the key to fully understanding the explosion mechanisms of the faint end of
the `normal' SNe on the WLR.  

We find evidence for differences in parameters for SNe\,Ia from different host
galaxies. SNe\,Ia in passive galaxies are more likely to be peculiar than those
in S-F galaxies. These peculiarities could be central in the debate on different
progenitor systems or populations. We found that there are three main
populations of SNe\,Ia: normal SNe\,Ia in S-F galaxies,  normal SNe\,Ia in
passive galaxies and sub-luminous SNe\,Ia. Normal ($M_{V}<-18\,mag$) SNe from
passive galaxies have a median $M_{V}$=--18.86$\pm$0.12\,mag, while those from
S-F galaxies have a median $M_{V}$=--19.25$\pm$0.06\,mag. Finally, we find that
$(15\pm10)\%$ of `normal' SNe\,Ia from S-F galaxies are likely to come from old
stellar systems. 

SNe\,Ia are, although ``standardisable", not an extremely homogeneous group of
objects. There are unusual events which could possibly result from multiple
progenitor scenarios. We have confirmed previous results  that sub-luminous SNe
favour passive galaxies, and have raised questions about the nature of `normal'
SNe\,Ia from passive galaxies. The key to examining these differences further
could be through finding a convincing method to calculate host galaxy extinction
on unusual SNe\,Ia, and with this it will be possible to put tighter constraints
on the parameter space SNe\,Ia can fill.    

\section{Acknowledgements} 
The authors would like to thank Mark Sullivan and Chris Usher for advice on how to treat the data.  The authors would also like to thank the anonymous referee for helpful comments and advice. The authors would like to thank the SNDB team for releasing their data to the public. This research has made use of the NASA/IPAC Extragalactic Database (NED) which is operated by the Jet Propulsion Laboratory, California Institute of Technology, under contract with the National Aeronautics and Space Administration. This research has made use of the CfA Supernova Archive, which is funded in part by the National Science Foundation through grant AST 0907903.

\appendix
\section{}

Table \ref{table:ref1} presents the references of the SNe which have come from individual papers.	

\begin{table}
  \centering
 \caption{The reference of the 11 SNe\,Ia mentioned in Table 1.}
 \begin{tabular}{cc}

 \hline
 SNe name & Reference \\
 \hline
 SN 1986G&\citet{SN1986G}\\
 SN 1990N&\citet{SN1990N}\\
 SN 1991bg&\citet{Krisciunas04}\\
 SN 1991T&\citet{Krisciunas04}\\
 SN 1998aq&\citet{Riess05}\\
 SN 2000ca&\citet{Krisciunas2004}\\
 SN 2000E&\citet{SN2000E}\\
 SN 2001bt&\citet{Krisciunas04}\\
 SN 2001cz&\citet{Krisciunas04}\\
 SN 2001el&\citet{sn2001el}\\
 SN 2003du&\citet{SN2003du}\\
 \hline
 \end{tabular}
\label{table:ref1}
\end{table}

\section{Quality of LC spine fit}
\begin{figure}
\centering
\includegraphics[scale=0.4]{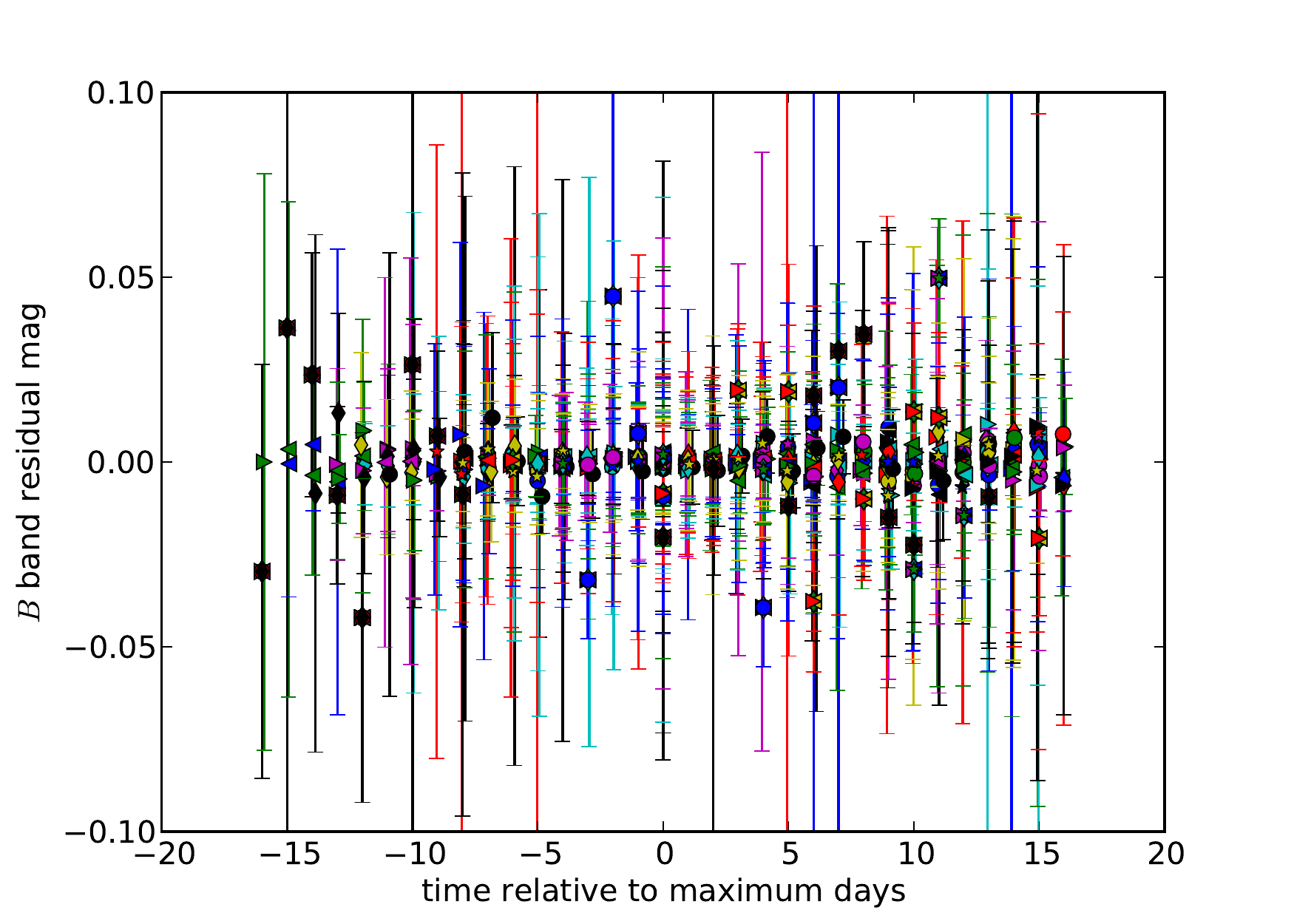}
\caption{The $B$ band residuals from the SN photmetry and LC spline fit, as a function on time. Data from 46 randomly selected SNe are used. }
\label{fig:Bresiduals}
\end{figure}

\begin{figure}
\centering
\includegraphics[scale=0.4]{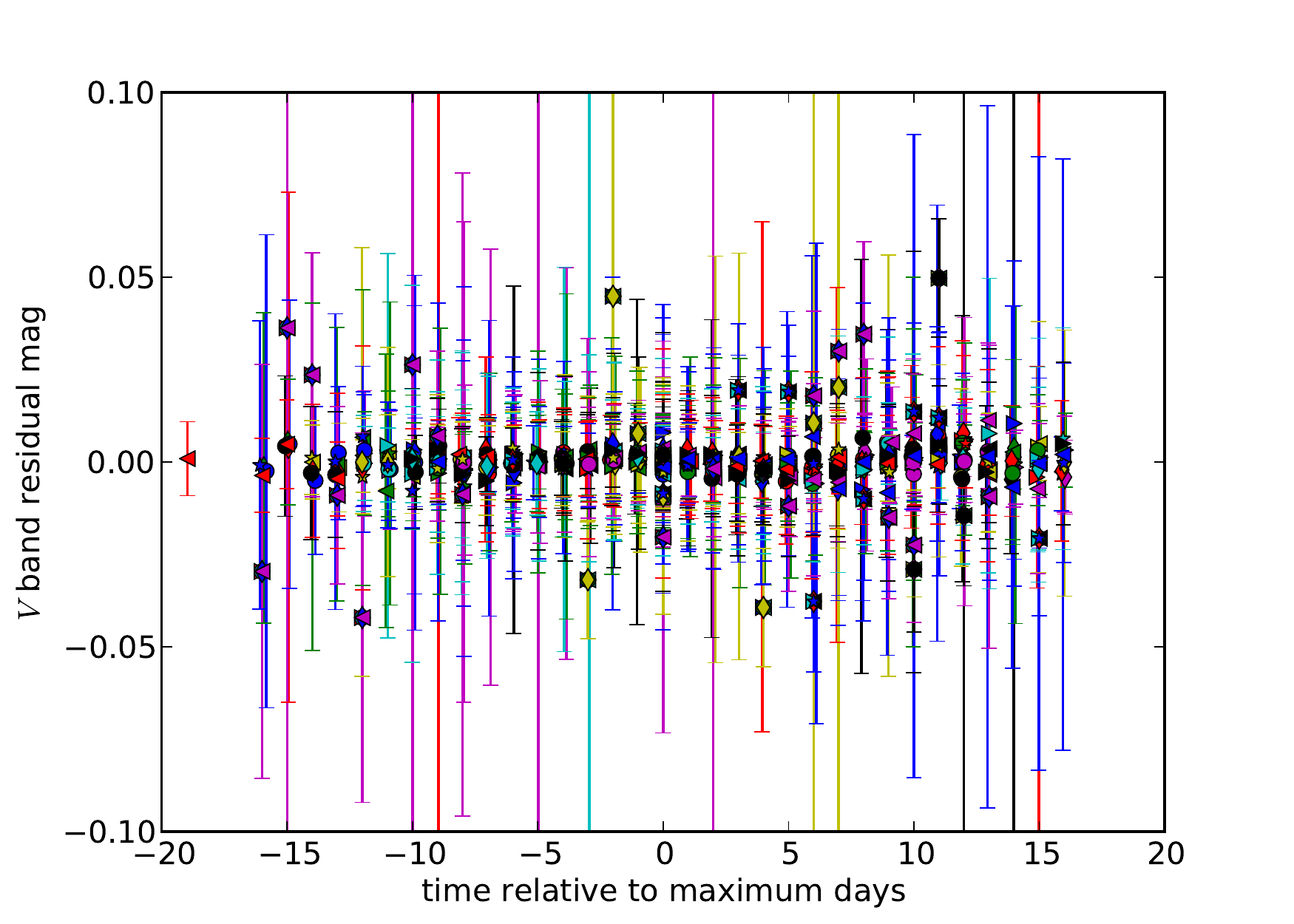}
\caption{The $V$ band residuals from the SN photmetry and LC spline fit, as a function on time. Data from 46 randomly selected SNe are used.}
\label{fig:Vresiduals}
\end{figure}

In this section we quantify the quality of our spline LC fits. In Figure \ref{fig:Bresiduals} we have plotted, in the $B$ band, the residuals between the SN data and fitted LC for 46 randomly selected SN from the sample. The residuals in this plot are less than 0.05\,mag for each SN. These are significantly smaller than the errors, which were taken as the errors on the photometric data. We can produce such small residuals as we only fit the LC until 20 days past maximum, and only SN with very good temporal coverage are selected. Figure \ref{fig:Vresiduals} shows the same residuals using the $V$ band, the residuals for these bands are also less than $0.05$\,mag. The residuals in these plots are constant as a function of time, therefore we have no systematic residuals with phase. We have not plotted the residuals for the errors as a MC approach has been used.

\end{document}